\documentclass[letter,twocolumn]{jpsj5}
\usepackage{graphicx}% Include figure files
\usepackage{dcolumn}% Align table columns on decimal point
\usepackage{amsmath,amsbsy,amssymb,graphics}
\usepackage{bm}% bold math
\usepackage{mathrsfs,color}

\begin{document}

%\preprint{APS/123-QED}

\title{
Composite electronic orders induced by orbital Kondo effect
%Kondo effect with non-Kramers configurations 
}

\author{Yoshio Kuramoto}

%\affiliation{
\inst{
Institute of Materials Structure Science, High Energy Accelerator Research Organization, \\
Tsukuba 305-0801, Japan
}

\date{\today}

%\begin{abstract}
\abst{
In a large number of rare-earth and actinide systems, Kondo effect 
tends to suppress magnetic order by making the spin singlet between localized and conduction electron spins.  
In the presence of orbital degrees of freedom, however, there emerge exotic electronic orders {\it induced} by Kondo effect.  
The orbital Kondo effect can collectively make diagonal and off-diagonal (superconducting) orders.  
With the particle-hole symmetry in conduction bands, these orders are all degenerate, forming a macroscopic SO(5) multiplet.
This paper discusses recent theoretical development on these electronic orders which are
relevant to Pr$^{3+}$ and U$^{4+}$ systems 
with even number of $f$ electrons per site.
In the superconducting order, each conduction-electron pair is coupled with local degrees of freedom, forming a composite entity with a staggered spatial pattern. 
The quasi-particle spectrum is best interpreted as virtual hybridization with resonant states at the Fermi level. 
Possible order parameter for URu$_2$Si$_2$ in the hidden order state
is discussed in the context of composite orders.
Briefly discussed are related issues such as homogeneous odd-frequency pairing and SO(5) theory for high-temperature superconductors.
}%\end{abstract}

\maketitle

\newcommand{\diff}{\mathrm{d}}
\newcommand{\imag}{\mathrm{Im}\,}
\newcommand{\imu}{\mathrm{i}}
\newcommand{\epn}{\mathrm{e}}

%main text

\section{Introduction}

Ever since the original paper by J. Kondo in 1964 \cite{kondo64}, physics including the Kondo effect has deepened and widened continuously.  The target systems have been extending from the original magnetic impurity systems to such large varieties as heavy electrons \cite{hewson93,kuramoto00}, semiconductor nanostructures\cite{goldhaber98}, metallic superlattices \cite{matsuda12}, and even high-density quark matters described by QCD \cite{ozaki15}.
The new directions can be divided into two parts: 
(i) impurity systems with controllable environment such as semiconductor nanostructures, and 
(ii) lattice of Kondo centers which interact mutually and give rise to collective phenomenon such as superconductivity.
The focus of this review is on the second category.  In particular, we are motivated by 
the idea that Kondo effect may work for {\it creating} a new kind of electronic order, in contrast to suppressing magnetic order as is customary perceived.

The key ingredient for Kondo-induced order is the orbital degrees of freedom \cite{cox87}.  Suppose that an impurity in the metallic matrix has the orbital (non-Kramers) doublet as a result of the crystalline electric field, and has an orbital exchange interaction with conduction electrons.
If the sign of the exchange is positive, which favors an orbital singlet, the orbital Kondo effect should work.
Obviously conduction electrons also have the spin degrees of freedom, which work as multiple (two) channels for orbital screening.  
As a result, an overscreening occurs and the orbital singlet cannot be formed.  
The idea of overscreening was first put forward by Nozi\`{e}res and Blandin \cite{nozieres80} in a different context. Namely, they
considered a situation where the impurity has only the spin degrees of freedom, but conduction electrons have both spin and orbital degeneracy.  
In this case the overscreening occurs for the spin degrees of freedom.
However, it is unlikely that each momentum of conduction bands has the orbital degeneracy.  
Hence the original setting does not seem to be realistic in actual systems.

With the overscreening of any form, finite entropy remains at each local site.  Then the system of orbital Kondo centers inevitably undergoes an electronic order to remove the entropy.
The simplest way, occurring even without Kondo effect, is forming an orbital (quadrupole or hexadecapole) order.  The resulting lower symmetry removes the orbital degeneracy, and the entropy vanishes.
There has been an enormous number of experimental and theoretical studies in this direction \cite{kuramoto09,santini09}.
With strong Kondo effect, however, other new kinds of order may occur.
For example, orbital singlets may form together with
spontaneous breakdown of the spin degeneracy.  
The resultant state breaks the time reversal, but
may not have the macroscopic magnetic moment \cite{nourafkan08,hoshino11,hoshino13}.
These exotic orders 
will be explained in detail in this paper.  

There is a long history in study of the overscreened Kondo effect.  
Theoretical status up to 1998 has been summarized in the extensive review by 
Cox and Zawadowski \cite{cox98}.
On the experimental side, however, the orbital Kondo effect still awaits unambiguous identification.  
Some of the promising candidates will be discussed later \cite{ott83,matsubayashi12,tsujimoto14,shimizu15}. 
Beginning with introductory description for the orbital Kondo effect, 
we shall review theoretical development achieved mainly after Ref.\citen{cox98}.
Our view is basically consistent with Ref.\citen{cox98} for the two-channel Kondo effect.
Concerning the transition to the novel ordered phases, however, we put forth a different viewpoint on the basis of recent progress.
It has been recognized \cite{emery92,abrahams93} that the composite order can  be viewed as an odd-frequency order of conduction electrons.  This view makes it practical to derive the transition temperature numerically, and resolves some confusion in the literature \cite{cox98,jarrell97}.

In the following section, we start with the impurity system where 
the orbital degrees of freedom of localized $f$ electrons are interacting with conduction electrons.
We give brief overview of the multi-channel {\it impurity} Kondo effect where 
the overscreening occurs.  We emphasize that dominant fluctuations from the ground state are composite objects involving both local and itinerant electrons, and that these objects are related to each other by a hidden SO(5) symmetry.
In the rest of the paper we deal with the main target; composite diagonal and off-diagonal orders induced in the orbital Kondo {\it lattice}.
Section \ref{sec:diagonal-order} discusses the diagonal composite order, while 
Section \ref{sec:off-diagonal} considers the superconducting composite order.
If the particle-hole symmetry is present in the conduction bands, both orders are related to each other by the SO(5) symmetry.  This aspect becomes most transparent in terms of fictitious hybridization of conduction electrons
with a resonant level at the Fermi level,
which is the subject of Section \ref{sec:hybridization}.
Possible relevance of the composite orders to actual systems are discussed in Section \ref{sec:experiment}.
Finally Section \ref{sec:discussion} discusses related subjects concerning odd-frequency superconductivity and SO(5) symmetry in single-band models.
We close the paper by summarizing and giving some outlook on the subject.

This paper is written mainly for non-experts of Kondo effect, although basic knowledge of interacting electrons is assumed.  Hence the reference list is by no means exhaustive. 
If appropriate review paper is available on the topic, many of original papers are not cited. 

\section{Non-Kramers Kondo impurity}
\label{sec:impurity}
\subsection{Orbital pseudo-spin and exchange interaction}
\begin{figure}[b]
\begin{center}
\includegraphics[width=70mm]{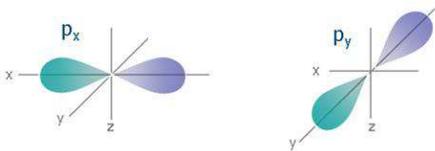}
\caption{
 Example of the simplest degenerate orbital states of two electrons with $p_x^2$ and $p_y^2$ configurations.
}
\label{pxpy}
\end{center}
\end{figure}

The simplest example of the orbital degrees of freedom is illustrated in Fig.\ref{pxpy} where two electrons of up and down spins occupy either $p_x$ or $p_y$ orbital.
In each orbital, two electrons form the spin singlet.   The two orbital states are conveniently expressed in terms of pseudo-spin with 
$|+\rangle$ for the $p_x^2$ state and
$|-\rangle$ for the $p_y^2$ state.  With the tetragonal symmetry around the $z$-axis, the states $|\pm\rangle$ are degenerate.

Analogous situation occurs in actual $f$-electron systems with $f^2$ configuration under the crystalline electric field (CEF).  However, the strong spin-orbit interaction makes the constituent $f$-electron states much more complicated.  
For example, the cubic CEF has doubly degenerate eigenstates labeled $\Gamma_3$, where the 
$f$-electron states have angular charge distributions such as
$x^2-y^2$ and $3z^2-1$ with $x^2+y^2+z^2=1$.
These charge distributions are regarded as orbitals.  
The operator describing the flipping of the orbital (or pseudo-spin) states 
$a,b$ 
is written as
$X_{ab} =|a\rangle\langle b|$. 
Then
the simplest local interaction between the $\Gamma_3$ states and conduction electrons is parameterized by the orbital exchange $J$, and is written as
\begin{align}
{\cal H}_{\rm ex} = \frac J2 \sum_{ab}X_{ab}\sum_\sigma c^\dagger_{b\sigma}c_{a\sigma},
\label{Hex}
\end{align}
where conduction electrons are characterized not only by the orbital $a, b$ but by the spin states $\sigma$.  
It is obvious that ${\cal H}_{\rm ex}$ is invariant under the point group operation.
Precisely speaking, $\sigma$ describes one of a time reversal (Kramers) doublet, which also has an orbital component with spin-orbit interaction.  For simplicity, however, we refer to $\sigma$ as "real spin".
We use Latin characters such as $a,b$ for indices of pseudo-spins, and Greek characters such as $\alpha,\beta,\sigma,\rho$ for real spins. 

The localized states with even number of $f$ electrons are called non-Kramers configurations.  Pr$^{3+}$ and U$^{4+}$ with two $f$-electrons belong to such cases.
Here
the Kramers theorem about the time-reversal degeneracy does not apply, and
the CEF ground state of a non-Kramers ion can either be a singlet, or a multiplet.
Those systems such as PrAg$_2$In \cite{yatskar96}, 
PrMg$_3$ \cite{tanida06}, 
PrPb$_3$\cite{onimaru04} and PrTi$_2$Al$_{20}$\cite{sakai11} 
have doubly degenerate states as the CEF ground state.
Since spatial distribution of wave functions in the doubly degenerate state
are different from each other, an electric multipole such as quadrupole and hexadecapole should arise in the ordered phase.
In addition, an imaginary coefficient in 
linear combination of wave functions brings about
magnetic degree of freedom such as octupoles ($2^3$) and triakontadipoles ($2^5$).

There are two main sources as the origin of $J$: 
(i) Coulomb repulsion between $f$ and conduction electrons, and 
(ii) hybridization between them \cite{cox98}.  These sources are analogous to those for the spin exchange \cite{kondo62} where the Coulomb interaction favors the ferromagnetic interaction, while the hybridization favors the opposite sign \cite{schrieffer66}.  Unfortunately, we are not aware of systematic study about the sign and magnitude of the orbital exchange $J$.  We assume $J$ positive, which is necessary to have renormalization to strong coupling.  
If we have $J<0$, the higher-order effect drives the system to decoupled $f$ and conduction electrons \cite{kondo64}.
Experimentally, there are many non-Kramers systems where $f$ electrons are localized well.  It is likely that these systems have negative $J$.  On the other hand, we expect $J>0$ for some U and Pr systems such as UBe$_{13}$ \cite{shimizu15}
and PrV$_2$Al$_{10}$ \cite{tsujimoto14}, which
show behaviors analogous to those of canonical Kondo systems such as CeCu$_2$Si$_2$ and CeB$_6$ \cite{kuramoto09}.  
In general, $J$ tends to be positive and large as itinerant character of $f$ electrons becomes stronger.

In terms of the pseudo-spin operator $\hat{\bm S}$ for $f$ electrons, the orbital permutation in Eq.(\ref{Hex}) is written as
\begin{align}
\sum_{ab\sigma}X_{ab}c^\dagger_{b\sigma}c_{a\sigma}
= \sum_{ab\sigma} \left( 
\hat{\bm S}_{ab}\cdot c^\dagger_{b\sigma}\bm\sigma_{ba} c_{a\sigma}
+\frac 12 \delta_{ab} c^\dagger_{b \sigma} c_{a\sigma}
\right),
\end{align}
where $c_{a \sigma}$ is the annihilation operator of the conduction electron at the impurity site with orbital 
$a =1,2$ (or written as $\pm$ if convenient) and spin $\sigma = \uparrow, \downarrow$.  The second term in the right-hand side (RHS) describes the potential scattering and will be neglected in the following.
By adding the kinetic energy of conduction electrons we obtain the Hamiltonian of the two-channel Kondo impurity:
\begin{align}
{\cal H}_{\rm KI} &= \sum _{\bm{k} a \sigma}  \varepsilon _{\bm{k}}  c_{\bm{k}a \sigma}^\dagger c_{\bm{k}a \sigma}
+ J 
\hat{\bm S} \cdot ( \hat{\bm{s}}_{\uparrow} + \hat{\bm{s}}_{\downarrow} )
, \label{eqn_2ch_KI}
\end{align}
where
$\hat{\bm{s}}_\sigma=\frac{1}{2}\sum_{ab} c^\dagger_{a\sigma} \bm{\sigma}_{ab}c_{b\sigma}$ is the pseudo-spin operator of conduction electrons with channel (real spin) components $\sigma$.
The Hamiltonian (\ref{eqn_2ch_KI}) has the SU(2) symmetry in both spin 
and orbital degrees of freedom, which are written as 
SU(2)$_{\rm S} \otimes$SU(2)$_{\rm O}$.

Let us mention briefly other models of non-Kramers CEF states.
If the CEF ground state is a singlet with small splitting $\Delta$ below a triplet, the spin Kondo effect occurs even for non-Kramers systems \cite{koga96,yotsuhashi02,otsuki05,hoshino09}.  
In the simplest picture, 
each electron of $f^2$ configuration has the antiferromagnetic interaction $J_2$ with conduction electrons, and competes with $\Delta$.
This situation is analogous to the two-impurity Kondo problem where the intersite  interaction, often called the RKKY interaction,  plays the role of $\Delta$.  
It is known that the two kinds of singlets, CEF and Kondo, crosses over in general as $J_2/\Delta$ increases from zero.  If there is a higher symmetry, however, there appears a quantum critical point where the ground state shares the same feature as that of the two-channel Kondo model \cite{jones88,affleck92}.
It has been argued \cite{kuramoto09,hoshino10} that
a strange order in PrFe$_4$P$_{12}$ is a consequence of the competition between $J_2$ and $\Delta$. 
Recently another new model has been proposed \cite{kusunose16} that takes the {\it j-j} coupling scheme to construct the $f^2$ CEF states.
Depending on the strength of $J_2$ and the orbital exchange, either orbital or spin Kondo effect dominates over the other.
Since we prefer a reasonable size of the review, the singlet CEF ground state 
will not further be discussed.

\subsection{Orbital Kondo Effect}
\begin{figure}[b]
\begin{center}
\includegraphics[width=0.85\linewidth]{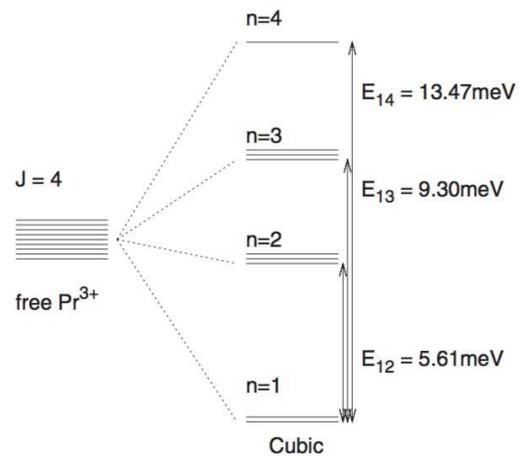}
\end{center}
\caption{
The CEF level structure determined by inelastic neutron scattering 
in PrTi$_2$Al$_{20}$.  The ground state ($n=1$) is the CEF doublet called $\Gamma_3$, while the first excited state ($n=2$) is the $\Gamma_5$ triplet
\cite{sato12}.
}
\label{PrTiAl-CEF}
\end{figure}

We prefer the term "orbital Kondo effect" rather than the "quadrupole Kondo effect" originally proposed \cite{cox87}, since the relevant multipole may be hexadecapole depending on the CEF states \cite{haule09,toth11}.
Both spin and orbital types of Kondo effect bring about 
logarithmic temperature dependence $\ln T$ of the electrical resistivity $\rho(T)$ at  temperatures higher than the characteristic 
temperature $T_K$, which is called Kondo temperature and gives the energy scale of the system.
However,  convincing examples of $\ln T$ behavior caused by the orbital Kondo effect are still lacking.
For example the CEF doublet $\Gamma_3$ system
PrTi$_2$Al$_{10}$ 
shows $\ln T$ behavior in $\rho (T)$ with maximum at $T\simeq 55$K as will be shown in 
Fig.\ref{rho(T)-exp} later \cite{sakai11,matsubayashi12}.
The CEF level structure has been determined by neutron scattering as shown in Fig.\ref{PrTiAl-CEF} \cite{sato12}.
Since the thermal population of first excited states ($n=2$) cannot be neglected for the temperature range $T>60$K, the origin of the $\ln T$ behavior may include  magnetic exchange.  
More famous candidate of the orbital Kondo effect is
UBe$_{13}$ where $\rho(T)$ increases with decreasing $T$, and saturates around $T\sim 40$ K\cite{ott83}, which is in sharp contrast with ordinary metals, and is often referred to as ``non-Fermi liquid". 
Superconductivity emerges below about 1 K from this non-Fermi liquid state \cite{ott83} as will be shown in Fig.\ref{rho(T)-exp} \cite{shimizu15}.  It is not trivial whether one can use the CEF picture in this material since the neutron scattering has probed only a broad feature \cite{shapiro85}.

On the other hand, Kondo effect is not seen in resistivity in such systems as 
 PrAg$_2$In \cite{yatskar96} and PrMg$_3$ \cite{tanida06}, which neither
show any symptom of electronic order at least down to 0.1  K.
As a result, the $T$-coefficient of the specific heat becomes huge, reaching to several J/(mole$\cdot$K$^2)$.
One of possible reasons for the absence of phase transition is splitting of the CEF doublet by crystalline disorder, which seems difficult to control in the Heusler structure \cite{yatskar96,tanida06}
.
Another reason may be that the orbital exchange $J$ is negative, leading to
ferro-coupling of $f$ and $c$ quadruples.
Unfortunately, there has been no detailed study to evaluate $J$ for non-Kramers doublet systems taking realistic electronic structure. 
We concentrate from now on to the case $J>0$.

\subsection{Nontrivial fixed point and perturbations}

Let us briefly discuss the scaling theory and the non-trivial fixed point for the model ${\cal H}_{\rm KI}$.
Following the elegant idea of Nozi\`{e}res and Blandin \cite{nozieres80}, we consider the hypothetical case where the number of screening channels is an arbitrary integer $n$.  The dimensionless coupling constant $g\equiv J_{\rm eff}\rho_c$,
 with $\rho_c=1/(2D)$ being the density of states of conduction band,
  obeys the renormalization group (RG) (or scaling) equation \cite{nozieres80}:
\begin{align}
\frac{dg}{dl} = -g^2 +\frac n2 g^3 +O(n g^4, n^2g^5) 
\label{3rdRG}
\end{align}
with $\ell \equiv \ln (D_{\rm eff}/D)$.  
The simplest method to derive Eq.(\ref{3rdRG}) will be the effective Hamiltonian formalism as detailed e.g. in Refs.\citen{kuramoto00,kuramoto98}.
The scaling means roughly that the set
$(J_{\rm eff}, D_{\rm eff})$ gives the same low-energy physics as the combination of the bare quantities $(J,D)$.  In other words, $J_{\rm eff}$ represents the $t$-matrix of a conduction electron with energy $D_{\rm eff}$\cite{anderson70}.

The fixed point of the RG corresponds to zero of the RHS.  
Note that the third-order term has a factor $n$ because of participation of all screening channels.
If we have a large value of $n$, which is actually unrealistic (but turns out useful), we may neglect the $O(g^4)$ terms near the fixed point.
We obtain $g=g_c$ at the fixed point as
\begin{align}
g_c = 2/n,
\end{align}
which is much smaller than unity with $n\gg 1$.
Then the neglected $O(g^4)$ terms are in fact smaller than the first two terms in the RHS.  

\begin{figure}
\begin{center}
\includegraphics[width=0.95\linewidth]{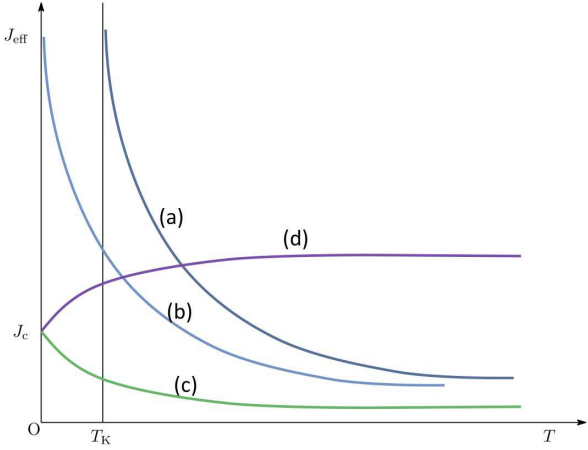}
\end{center}
\caption{
Temperature-dependence of the effective interaction: (a) lowest-order renormalization, (b) correct behavior in the ordinary Kondo model, (c) third-order renormalization, (d) behavior with the large bare coupling $g>g_c$.
In the ordinary Kondo model ($n=1$), (c) and (d) are wrong, but are qualitatively correct for  
$n \ge 2$.
}
\label{J_eff(T)}
\end{figure}

Remarkably, this argument in Ref.\citen{nozieres80} remains qualitatively valid down to the realistic case of $n=2$.  The essence of the validity comes from the physical phenomenon of ``overscreening" that is common to all $n\ (>1)$.  For example, if the local pseudo-spin makes a singlet with a conduction pseudo-spin with channel $\sigma$, another conduction channel $\bar{\sigma}$ still keeps the active orbital pseudo-spin.  Hence the entropy of the system cannot go to zero.
Figure \ref{J_eff(T)} illustrates the effective coupling as a function of temperature $T$ which corresponds to $D_{\rm eff}$.
Together with the renormalization to the non-trivial fixed point, we also show the case of the ordinary Kondo model including the renormalization at lowest order.

The non-trivial fixed point becomes unstable once the SU(2)$_{\rm S} \otimes$SU(2)$_{\rm O}$ symmetry for spin and orbital spaces is broken by an external perturbation. 
For example, 
suppose that the orbital degeneracy of the local pseudo-spin is slightly broken
by uniaxial pressure, which can be represented as a pseudo-Zeeman term 
$h_z \hat{S}_z$.  
Then the non-trivial ground state will change into the decoupled CEF singlet and the Fermi sea of conduction electrons with no remaining entropy.  
The effective coupling $g$ renormalizes to zero below the temperature corresponding to the splitting.

On the contrary,  
if the orbital exchange becomes larger for a channel than another,  the effective coupling for that channel goes to infinity, and the other goes to zero.
The ground state is a local Fermi liquid without residual entropy.
Thus these important (relevant) perturbations are represented as
\cite{affleck92}:
\begin{align}
V_1^z = h_1\hat{S}^z, \quad
V_2^z = h_2\hat{\bm S}\cdot (\hat{\bm{s}}_\uparrow-\hat{\bm{s}}_\downarrow),
\end{align}
where $V_1^z$ has the pseudo magnetic field $h_1$ that breaks the non-Kramers degeneracy, and $V_2^z$ breaks the equivalence of two spin channels.  
Because of the SU(2)$_{\rm S}$ symmetry, we have
also $x,y$ components for $V_1, V_2$ as will be shown shortly.
Note that $V_1^z$ involves only localized electrons, while $V_2^z$ is a composite object involving both local and conduction parts.  Hence $V_2$ is more subtle and nontrivial.

We now focus on $V_2^z$ and regard $\hat{\bm S}\cdot (\hat{\bm{s}}_\uparrow-\hat{\bm{s}}_\downarrow)$
as the $z$-component of a vector operator $\bm\Psi$.  
In general the $\mu$-component $\Psi^\mu$ with $\mu=x,y,z$ is given by
\begin{align}
\Psi^\mu= \frac 12
\hat{\bm S}\cdot 
{\bm\sigma}_{ab}
c_{a\sigma}^\dagger 
\sigma^\mu_{\sigma\rho}
c_{b\rho}
\label{Psi}
\end{align}
Here and in the rest of the present paper, 
we take the Einstein convention, if obvious, to omit summation symbols over repeated spin and orbital variables.
It is convenient to introduce the combinations
$\Psi^\pm =\Psi^x\pm i\Psi^y$ giving
\begin{align}
\Psi^+=
\hat{\bm S}\cdot {\bm \sigma}_{ab}
c^\dagger_{a\uparrow}c_{b\downarrow},
\quad
\Psi^-=
\hat{\bm S}\cdot {\bm \sigma}_{ab}
c^\dagger_{a\downarrow}c_{b\uparrow}, 
\end{align}
which has flips of both real ($\uparrow,\downarrow$) and pseudo  ($a,b$) 
spins. 

Let us consider the case where the degenerate conduction bands have a
particle-hole (PH) symmetry, which is
realized if the degenerate conduction bands
have flat density of states $\rho_c$ between the cut-offs $\pm D$, and the Fermi level is at the center of the band, i.e. $\mu =0$.
With the PH symmetry, 
$\cal H_{\rm KI}$ in Eq.(\ref{eqn_2ch_KI}) commutes with corresponding generators.  As an example we consider $P_\sigma$ for each spin (channel) $\sigma$ defined by
\begin{align}
P_\sigma c_{a\sigma} P_\sigma^{-1} = \epsilon_{ab}  c^\dagger_{b\sigma}, \quad 
P_\sigma c_{a\sigma}^\dagger P_\sigma^{-1} = \epsilon_{ab}  c_{b\sigma}, \quad 
\label{P_sigma}
\end{align}
where $\epsilon = i\sigma^y$ is the antisymmetric unit tensor.
For $c_{a\rho}$ with $\rho\neq \sigma$,
$P_\sigma$ behaves as the identity operator.  It follows then
\begin{align}
P_\sigma c_{a\sigma}^\dagger {\bm \sigma}_{ab} c_{b\sigma}P_\sigma^{-1} = 
c_{a\sigma}^\dagger {\bm \sigma}_{ab} c_{b\sigma},
\end{align}
without summation over $\sigma$.  Namely 
the PH transformation conserves the pseudo-spin.
Since the interaction in $\cal H_{\rm KI}$ depends only on the pseudo-spin,  we obtain 
$[{\cal H}_{\rm KI}, P_\sigma ]= 0$ under the PH symmetry.

By straightforward calculation we obtain
\begin{align}
P_\uparrow \Psi^+ P_\uparrow^{-1} 
&= \frac 12 
\hat{\bm S}\cdot (\epsilon{\bm \sigma})_{ab}\epsilon_{\sigma\rho}
c_{a\sigma}c_{b\rho} 
\equiv \Phi 
\equiv \hat{\bm S}\cdot {\bm t}
, \\
P_\downarrow \Psi^+ P_\downarrow^{-1} 
&= \frac 12 
\hat{\bm S}\cdot ({\bm \sigma\epsilon})_{ab}\epsilon_{\sigma\rho}
c^\dagger_{a\sigma}c^\dagger_{b\rho} 
=\Phi^\dagger
= \hat{\bm S}\cdot {\bm t}^\dagger
,  
\label{Phi}
\end{align}
where the quantities $\Phi$ and $\bm t$ are introduced.
The operator $\bm t$ annihilates a pair of conduction electrons
with channel singlet (Cs) and pseudo-spin triplet (St).
Other combinations of $P_\sigma$ and $\Psi^-$ give either $\Phi$ or $\Phi^\dagger$.   
Here, 
following the literature \cite{jarrell97,anders02},
we use the labels C and S indicating `channel' and `spin', 
and s and t indicating `singlet' and `triplet'.
This terminology is unfortunately confusing here since `spin' means actually pseudo-spin (orbital) in our case, and `channel' represents the real spin.
With this caveat, we follow the conventional terms for easy comparison with the literature.

With $P_\sigma$ commuting with ${\cal H}_{\rm KI}$,
Hermitian operators $\Psi^\mu \ (\mu =x,y,z)$ and 
$\Phi_{\rm R}=(\Phi+\Phi^\dagger)/2, \Phi_{\rm I}=i^{-1}(\Phi-\Phi^\dagger)/2 
$ have the same fluctuation spectrum.
The quintet of operators thus represent a symmetry larger than the obvious one:
SU(2)$_{\rm S} \otimes$SU(2)$_{\rm O}$.
Namely, by incorporating the PH symmetry, the system acquires the SO(5) symmetry or, equivalently, the Sp(4) symplectic symmetry \cite{affleck92}.
The operator $P_\sigma$ belongs to the set of Sp(4) generators.
We shall return to this problem when we discuss the symmetry aspect of order parameters in later sections.

\section{Composite Diagonal Order}
\label{sec:diagonal-order}

\subsection{Relation to odd-frequency order}

We now turn from the two-channel
Kondo impurity to the lattice (2chKL) as a model of $f$-electron systems with non-Kramers doublets. 
The Hamiltonian is given by
\cite{jarrell96}
\begin{align}
{\cal H}_{\rm KL} = 
\sum _{\bm{k} a\sigma} 
 \varepsilon _{\bm{k}} c_{\bm{k}a \sigma}^\dagger c_{\bm{k}a \sigma}
+ J \sum_{i\sigma} \hat{\bm S}_i \cdot \hat{\bm{s}}_{i\sigma }
, \label{eqn_2ch_KLM}
\end{align}
which is to be compared with ${\cal H}_{\rm KI}$.

In the lattice system, the residual entropy is removed by spontaneous breaking of the symmetry, namely by an electronic ordering.
The simplest order 
has the mean-field described by $V_1^z$. 
Since this kind of order is connected continuously with a simple multipole order of localized electrons, 
characteristics peculiar to two channels may not appear.
On the other hand, another order described by $V_2^z$ 
is more interesting since it
breaks the time-reversal symmetry 
even with zero magnetic moment: $\langle
\hat{\bm{s}}_\uparrow-\hat{\bm{s}}_\downarrow\rangle=0.$
In contrast with the ordinary Kondo lattice that can stay paramagnetic down to zero temperature, the 2chKL must have some kind of order because of the remaining entropy otherwise.

Let us first discuss symmetry breaking of channels, which usually corresponds to magnetic order of conduction electrons.
We introduce the following operator dependent on imaginary times:
\begin{align}
O_i^{\rm ch}(\tau,\tau') \equiv 
c_{ia\sigma} ^\dagger (\tau) \sigma_{\sigma\rho}^z
c_{ia\rho} (\tau')
.
 \end{align}
With a channel order, we obtain 
$\langle O_i^{\rm ch}(\tau,\tau)\rangle \neq 0$.
However, the channel symmetry is broken even if the equal-time average is zero, provided $\langle O_i^{\rm ch}(\tau,\tau') \rangle \neq 0$
for $\tau\neq \tau'$.
In particular, if the following quantity including the $\tau$-derivative:
\begin{align}
{\cal O} (\tau)\equiv 
\sum_i c_{ia\sigma}^\dagger (\tau) \sigma_{\sigma\rho}^z
\dot{c}_{ia\rho} (\tau)
,
\label{cal-O}
\end{align}
has a finite average, the resulting order  
is an example of odd-frequency (OF) orders, which will be discussed in more detail later for superconductivity.
Here we show that this order is equivalent to the homogeneous order of $\Psi^z$ defined by Eq.(\ref{Psi}).
Namely the commutator with ${\cal H}_{\rm KL}$ corresponding to the $\tau$-derivative gives
\begin{align}
{\cal O} = 
\sum _{\bm{k}} 
 \varepsilon _{\bm{k}} c_{\bm{k}a \sigma}^\dagger \sigma_{\sigma\rho}^z
 c_{\bm{k}a \rho}
+ J \Psi^z
,
\label{cal_O}
\end{align}
where we have redefined 
\begin{align}
\Psi^z \equiv \frac 12\sum_i
\hat{\bm S}_i \cdot 
{\bm\sigma}_{ab}
c_{ia\sigma}^\dagger 
\sigma^z_{\sigma\rho}
c_{ib\rho},
\end{align}
by including the site summation for the lattice system.

Let us consider how one can detect the composite order $\Psi^z$.  Obviously the most interesting is the experimental detection, which is discussed later.  In the numerical calculation for the 2chKL, the most powerful approach at present is the combination of the continuous-time quantum Monte Carlo (CT-QMC) \cite{gull11} and the dynamical mean-field theory (DMFT) \cite{georges96}.  The basic idea is to use the CT-QMC to solve the impurity problem in the effective medium, which is determined self-consistently by the DMFT \cite{hoshino13,hoshino-yk14}.  Here we do not go into details of numerical methods, which have been reviewed in detail \cite{gull11,georges96}.
It is possible to check whether $\Psi^z$ is finite or not at each temperature by direct evaluation.  
At the transition temperature, $\Psi^z$ begins to be finite and its fluctuation should diverge.
However, it is difficult to compute the corresponding response function 
since $\Psi^z$ is a composite object.

Since the $\tau$-derivative form of $\cal O$ includes only conduction electrons, its response function may be derived without including localized electrons explicitly.  
Aiming at deriving the response function, 
we work with the two-particle Green function:
\begin{align}
\chi^{\rm ch}_{ij} (\tau_1, \tau_2, \tau_3, \tau_4) = \langle T_\tau
O^{\rm ch}_i (\tau_1, \tau_2) O^{\rm ch}_j (\tau_3, \tau_4)
\rangle
,
\end{align}
where 
$T_\tau$ is the time-ordering operator. 
We perform the Fourier transform to Matsubara frequencies
$\varepsilon_n = (2n+1)\pi T$ as
\begin{align}
\chi^{\rm ch}_{\bm q} (\imu\varepsilon_n, \imu\varepsilon_{n'})
&= \frac{1}{N\beta^2} \sum_{ij}\int_0^\beta \hspace{-2mm} \diff \tau_1 \cdots \diff\tau_4
\ \chi^{\rm ch}_{ij} (\tau_1, \tau_2, \tau_3, \tau_4)
\nonumber \\
\times &
\epn^{-\imu \bm q \cdot (\bm R_i - \bm R_j)}
\epn^{\imu \varepsilon_n    (\tau_2 - \tau_1)}
\epn^{\imu \varepsilon_{n'} (\tau_4 - \tau_3)}
.
\end{align}
The response function of $\cal O$ is related to the derivative:
\begin{align}
\frac{\partial^2}{\partial\tau_2 \partial\tau_3}
\chi^{\rm ch}_{ij} (\tau_1, \tau_2, \tau_3, \tau_4). 
\label{tau_23-derivative}
\end{align}
The Fourier transform of Eq.(\ref{tau_23-derivative}) leads to
factors 
$\imu \varepsilon_n$ and $\imu \varepsilon_{n'}$.
The $\tau$-derivatives in Eq.(\ref{tau_23-derivative}) also give rise to a delta-function part due to the time-ordering.
In fact, one can prove the following relation \cite{hoshino11}:
\begin{eqnarray}
&&
\hspace{-10mm}
\frac 1N 
\int_0^\beta \langle {\cal O}(\tau){\cal O}^\dagger \rangle \diff \tau = 
\nonumber \\
&&
\hspace{-10mm}
- \frac{1}{\beta} \sum_{nn'}  \varepsilon_n  \varepsilon_{n'} \, 
\chi_{\bm q=0}^{\rm ch} (\imu \varepsilon_n, \imu \varepsilon_{n'}) e^{\imu \varepsilon_n 0^+}e^{\imu \varepsilon_{n'} 0^+}
\hspace{-1mm} + \hspace{-1mm} \frac 2N \langle {\cal H} \rangle
, \label{eq_sus2}
\end{eqnarray}
where convergence factors enter in frequency summations.
The second term in the RHS comes from delta-functions associated with the time-ordering.
Because of the second term, the sign of the first term is indefinite, even though the LHS is positive definite.
This point is crucial in interpreting the numerical results for the OF susceptibilities.

At the transition temperature to the ordering of $\cal O$, the LHS and the first term in the RHS of Eq.(\ref{eq_sus2}) diverge.
In the numerical calculation, the summation over Matsubara frequencies with convergence factors is very slow and awkward.  Therefore alternative scheme has been proposed \cite{jarrell97,anders02}.
Namely, one regards 
$\hat{\chi}_{nn'}\equiv \chi_{\bm q=0}^{\rm ch} (\imu \varepsilon_n, \imu \varepsilon_{n'})$
as a matrix in the space of Matsubara frequencies.  Then the divergence in Eq.(\ref{eq_sus2}) means that the maximum eigenvalue of $\hat{\chi}$ 
becomes infinity in the subspace of odd functions of $n$.
Let us consider then what happens if one replaces $\varepsilon_n$ by another odd function $g_n$ in Eq.(\ref{eq_sus2}).  
As long as the eigen vector with the maximum eigenvalue has a finite overlap with both $\varepsilon_n$ and $g_n$, the divergence occurs at the same temperature.
With this consideration 
one may choose $g_n = \tanh (\varepsilon_n/D)$ or simply $g_n={\rm sgn}(n)$.
These choices lead to much faster convergence in the frequency summation.

\begin{figure}[b]
\begin{center}
\includegraphics[width=85mm]{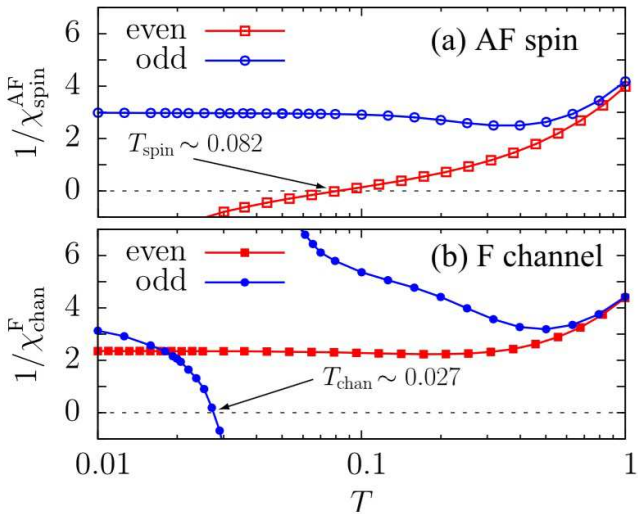}
\caption{
Temperature dependence of 
$1/\chi_{\rm even} $ and $1/\chi_{\rm odd}$ for half-filled ($n_c=2$) conduction bands
corresponding to (a) AF-pseudo-spin (orbital) and (b) F-channel (real spin) susceptibilities \cite{hoshino11}.
Phase transitions are signaled by the zero crossing of inverse susceptibilities.
}
\label{fig_suscep}
\end{center}
\end{figure}

Figure \ref{fig_suscep} shows the susceptibilities numerically derived at half-filling $n_c=2$ \cite{hoshino11}.  The staggered (AF) pseudo-spin susceptibility in panel (a) corresponds to the choice:
\begin{align}
O_i^{\rm orb}(\tau,\tau') \equiv 
c_{ia\sigma} ^\dagger (\tau) \sigma_{ab}^z
c_{ib\sigma} (\tau')
,
 \end{align}
which actually corresponds to orbital degrees of freedom.   
The even-frequency (EF) part diverges at $T\sim 0.082$ with the unit $D=1$ for the half-width of the conduction bands with semi-elliptic density of states.  The OF part remains finite for all temperatures.
The homogeneous (F) channel susceptibility shown in panel (b) does not show any indication of divergence for the EF part.  On the other hand, the OF part becomes negative as $T$ decreases, and then diverges from negative side at $T\sim 0.027$.
The divergence indicates the onset of long-range order with $\Psi^z$.
As we have remarked, the divergence is independent of the choice of $g_n$ although the magnitude of the susceptibility at higher temperature does depend on $g_n$.

\begin{figure}
\begin{center}
\includegraphics[width=85mm]{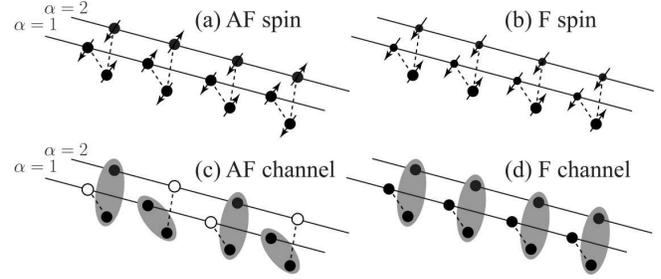}
\caption{
Schematic illustration of the ordered states in the 2chKL \cite{hoshino11}.
Here $\alpha$ represents the channel 1 or 2 (up or down of real spins), and 
the arrows show pseudo-spins. The shaded ovals show an orbital Kondo singlet, and
the open circles indicate the absence of conduction electrons.  The average occupation number $n_c$ per site is 2 in (a), (b), (d) and 1 in (c).
}
\label{fig_illust}
\end{center}
\end{figure}

Figure \ref{fig_illust} illustrates the ordering pattern with the conduction electron number $n_{\rm c} = 1,2$ per site \cite{hoshino11}.  Physically the F channel order with $\Psi^z$ in (d)
describes the breakdown of the spin symmetry;
the orbital Kondo insulator appears involving down-spin 
$\alpha = 2\ (\sigma=\downarrow)$ of conduction bands, while the up-spin ($\alpha=1$) conduction electrons make a Fermi liquid \cite{nourafkan08,hoshino11}.  The different spatial distributions of magnetic moments 
means the broken time-reversal, although there is no net magnetization.
In this sense, the order parameter $\Psi^z$ represents a state with itinerant magnetic multipoles.
If the non-Kramers degrees of freedom generates quadrupoles (hexadecapoles),  $\Psi^z$ represents an itinerant octupole (triakontadipole) order.
However, the orbital RKKY interaction makes the order (a) assigned as AF spin more stable than (d) with $n_{\rm c}=2$ \cite{jarrell97}.
In the non-Kramers system, (a) actually represents the AF orbital order. 
On the other hand,  the AF-channel order (c), which is antiferromagnetism of real spins, is most stabilized at $n_{\rm c} = 1$.  
The channel-symmetry breaking in (c) and (d) is a characteristic of the two-channel model, and cannot be explained by the RKKY interaction. Namely the Hamiltonian (\ref{eqn_2ch_KLM}) includes only the interaction between pseudo-spins rather than channels.

Surprisingly, the F-channel ($\Psi^z$) order is converted to 
a composite superconductivity by a unitary transformation.  
This is a consequence of the hidden
SO(5) symmetry as 
will be discussed in greater detail later.
The overall phase diagram of the 2chKL 
will be shown later in Fig.~\ref{fig_phase2}.

\subsection{Single-particle spectrum}

Now we consider the single-particle spectrum for the F-channel order, which leads us to the simple effective Hamiltonian at low energies.
We define the spectrum for the component $a,\sigma$ by
\begin{align}
A_{a\sigma}(\bm{k},\omega) = -\frac{1}{\pi}
\imag G_{a\sigma} (\bm{k}, \omega + \imu \eta)
, \label{eq_spect_def}
\end{align}
where $\eta = +0$.
The single-particle Green function is given by
\begin{align}
G_{a\sigma} (\bm{k}, z) ^{-1}
= z  - \varepsilon_{\bm{k}} - \Sigma _{a\sigma} (z)
, 
\end{align}
where $z$ is a complex energy and $\Sigma _{a\sigma}(z)$ is the self energy.
In the DMFT, the wave-vector dependence enters into the single-particle Green function only through $\varepsilon_{\bm{k}}$.

The spectrum in the paramagnetic state has a broad feature
(not shown),  
which is consistent with the incoherent metallic state \cite{jarrell96}.
The peak energy at each momentum is almost the same as the non-interacting one\cite{hoshino13}.
This non-Fermi liquid behavior originates from 
interaction of localized spins with overscreening degenerate channels.
In particular the Kondo singlet state cannot be formed.

\begin{figure}
\begin{center}
\includegraphics[width=50mm]{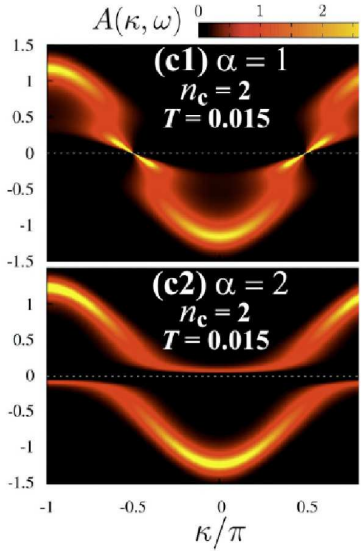}
\caption{
Spectra $A(\kappa,\omega)$ of conduction electrons with channel (a) $\alpha=1$ (spin up) and (b) $\alpha=2$ (spin down) \cite{hoshino13}.
The intensity is represented by different colors.  
See text for the definition of $\kappa$.
}
\label{spectrum_Psi}
\end{center}
\end{figure}

Figure \ref{spectrum_Psi} shows the numerical result \cite{hoshino11,hoshino13} in the ordered phase. 
The parameter $\kappa$ is introduced by $\varepsilon_{\bm{k}} = - D \cos \kappa$ so that the spectrum is visualized as if in a one-dimensional system.
The spectrum for channel (real spin) $\alpha=2$ is clearly 
characterized by a hybridization with localized levels precisely at the Fermi level.
On the other hand, the spectrum for $\alpha=1$ shows a Fermi liquid behavior with sharp Fermi surface.  The broadening becomes substantial away from the Fermi level.
Near the Fermi level, the Green function for both channels are parameterized as
\begin{align}
G_{a\sigma} (\bm{k}, z) 
&\sim \frac{a_\sigma}{z
- a_\sigma \varepsilon_{\bm{k}}  - a_\sigma{V_\sigma}^2/z} 
, \label{eq_FL_g}\\
\Sigma_{a\sigma} (z) 
&= \frac{{V_\sigma}^2}{z}
+ b_\sigma z + O\left( z^2\right)
, \label{eq_FL_sig}
\end{align}
where the renormalization factor is given by $a_\sigma = (1-b_\sigma)^{-1}$.
For $\alpha = 1\ (\sigma=\uparrow)$, numerical analysis of the self-energy 
\cite{hoshino13} gives
 $a_\uparrow \simeq 0.51$ and $V_\uparrow \simeq 0.00$ at $T=0.005$.
This means that the conduction electrons with spin up shows the Fermi-liquid behavior, and hybridization is absent.
For $\alpha = 2\ (\sigma=\downarrow)$, on the other hand, the values obtained are 
$a_\downarrow \simeq 0.43$, $V_\downarrow \simeq 0.33$.
This indicates the behavior of the Kondo insulator with
effective hybridization $V_\downarrow$.
Thus, the spectrum displays the admixture of the Fermi liquid and Kondo insulator.
From Fig.\ref{spectrum_Psi} it is apparent that hybridization gap in down spins causes lower kinetic energy than up spins.  Then we recognize that both constituents of the operator 
$\cal O$ in
Eq.(\ref{cal_O}) have non-zero average in the ordered phase.

Away from half filling, there emerges a doped Kondo insulator for $\sigma = \downarrow$  which behaves as heavy Fermi liquid at sufficiently low temperature.
The ordered phase in the doped case then consists of two different Fermi liquids dependent on channels.  

The effective hybridization picture,  which corresponds to the strong-coupling limit ($J\rightarrow \infty$),
has already been used for the standard Kondo lattice \cite{hewson93,kuramoto00}.  In the present case, the channel-symmetry breaking is essential to realize the effective hybridization.
Namely, 
in the paramagnetic state, we do not see any indication of hybridization in contrast to the ordinary Kondo lattice.

\section{Composite superconductivity}
\label{sec:off-diagonal}
\subsection{SO(5) symmetry at half-filling}

We have shown for the impurity system that leading fluctuations consist of a three-component vector $\bm\Psi$, Eq.(\ref{Psi}), and pairing fluctuations 
$\Phi, \Phi^\dagger$, Eq.(\ref{Phi}).
With the PH symmetry in half-filled conduction bands, these five fluctuations have exactly the same spectrum.
The underlying SO(5) symmetry remains effective also for the lattice system described by ${\cal H}_{\rm KL}$.
In particular a symmetry-broken ground state has the same energy with other four symmetry-broken states. 
Moreover, the excitation spectra are also common to these ordered phases as will be demonstrated in section \ref{sec:qp-spectra}.

Let us proceed to off-diagonal orders with particular attention to the symmetry aspect.
We assume the bipartite lattice with A, B sublattices, and introduce the site-dependent PH transformation:
\begin{align}
P_\downarrow c_{ia\downarrow} P_\downarrow^{-1} = \pm 
\epsilon_{ab}c_{ib\downarrow}^\dagger 
\label{PH_site}
\end{align}
where the sign factor is +1 for A and -1 for B sublattices.
Equivalently one may use $\exp({\rm i}\bm Q\cdot\bm R_i)=\pm 1$ 
 with $\bm Q$ corresponding to the staggered order.
Because of the phase factor, the hopping term between the nearest-neighbor sites remains the same under the PH transformation.  
The invariance can be illustrated by the hopping of spin-down electrons:
\begin{align}
t c_{{\rm A}a}^\dagger c_{{\rm B}a} 
& \rightarrow 
 -t c_{{\rm A}b} c_{{\rm B}b}^\dagger
= t c_{{\rm B}b}^\dagger c_{{\rm A}b} , 
\label{kinetic}
\end{align}
where the site indices are replaced by the sublattice indices A and B, and the spin indices are omitted.
The up-spin electrons are not affected by $P_\downarrow$.

The PH transformation $P_\downarrow$ for $\Psi^+$ gives
\begin{align}
P_\downarrow \Psi^+ P_\downarrow^{-1}
&= \frac 1{2}\sum_i e^{{\rm i}\bm Q\cdot \bm R_i}
\hat{\bm S}_i \cdot \left( 
{\bm \sigma} \epsilon\right)_{ab}\epsilon_{\sigma\rho}
c^\dagger_{ia\sigma}c^\dagger_{b\rho} 
\nonumber\\
&\equiv \Phi(\bm Q)^\dagger, 
\end{align}
which is to be compared with Eq.(\ref{Phi}) for the impurity system.
Here $\Phi(\bm Q)^\dagger $ represents a staggered and composite superconducting order.
In the same way, we can show 
\begin{align}
P_\downarrow  \Psi^-P_\downarrow ^{-1} =\Phi (\bm Q), \quad
P_\downarrow  \Psi^z P_\downarrow ^{-1} = \Psi^z .
\end{align}
On the other hand, explicit calculation shows
\begin{align}
P_\downarrow  \Phi (\bm Q) P_\downarrow ^{-1} =-\Psi^-, 
\end{align}
which also follows immediately from the definition given by Eq.(\ref{PH_site}).

\subsection{Strong-coupling limit}

It is instructive to take the strong coupling limit for visualizing the nature of composite order $\Phi$.  
Suppose
we have a doublet of either (i) pseudo-spin or (ii) channel at each site.
In each case, 
we can combine the doublets of two sites to make a singlet.  
For a finite-sized Hamiltonian, however, we cannot have the symmetry-broken eigenstate. 
Hence we shall deal with local eigenstates 
of the Hermitian operator $\Phi_{\rm R}= (\Phi+\Phi^\dagger)/2$. 

\begin{figure}
\begin{center}
\includegraphics[width=70mm]{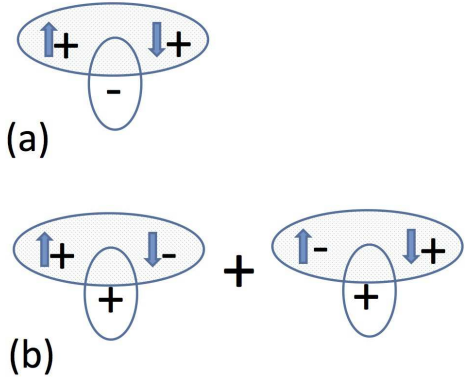}
\caption{
Illustration of the local composite pair where $\pm$ represents pseudo-spins, and arrows indicate channel (real spin) indices.
(a) $ |{\rm CsSt}_1\rangle \otimes |-\rangle$ and
(b) $ |{\rm CsSt}_0\rangle \otimes |+\rangle$.
}
\label{Phi_eigenstate}
\end{center}
\end{figure}

Let us begin with the case (i) for a single site.
We start from the state 
$|0\rangle\otimes |\pm\rangle \equiv |0\pm\rangle$ where 
$|0\rangle$ is a vacuum of conduction electrons, and $|\pm\rangle$ is the $f$-state doublet. Then we obtain
\begin{align}
\Phi^\dagger |0\rangle\otimes |+ \rangle &= 
2 |{\rm CsSt}_1\rangle \otimes |-\rangle +
 |{\rm CsSt}_0\rangle \otimes |+\rangle \nonumber\\
&\equiv \sqrt 5 |2+\rangle
,
\end{align}
where the pseudo-spin triplets of two electrons
with $z$-components 1,0
are given by
\begin{align}
 |{\rm CsSt}_1\rangle & = c_{+\uparrow}^\dagger c_{+\downarrow}^\dagger |0\rangle, \\
 |{\rm CsSt}_0\rangle & = \frac 1 2
\sigma^x_{ab}\epsilon_{\sigma\rho}
 c_{a\sigma}^\dagger c_{b\rho}^\dagger 
 |0\rangle, 
\end{align}
Figure \ref{Phi_eigenstate} illustrates these states.
On the other hand, starting from 
$|\tilde{0} \rangle\otimes |+\rangle \equiv |\tilde{0}+\rangle$ 
where 
$|\tilde{0} \rangle$ is the fully occupied state 
with $n_c=4$, we obtain the same state by
\begin{align}
\Phi |\tilde{0}+ \rangle = \sqrt 5 |2+\rangle = 
\Phi^\dagger |0 + \rangle .
\end{align}
Hence $\Phi_{\rm R}=(\Phi+\Phi^\dagger)/2$ is represented by a $3\times 3$ matrix with the basis set $|0+\rangle, |2+\rangle$ and $|\tilde{0}+\rangle$.
We can repeat the same analysis starting from $|0-\rangle$.
The eigenvalues $x_\mu$ and corresponding eigenvectors $\phi_\mu(a)$ 
with $\mu=0,\pm$ and $a=\pm$ are derived as
\begin{align}
 x_0&=0:\quad  \phi_0 (a) = \left( |0a \rangle+|\tilde{0}a  \rangle  \right)/{\sqrt 2}, \\ 
 x_\pm &=\sqrt{10}: \quad \phi_\pm (a) = |2+\rangle /\sqrt{2} \pm
\left( |0a \rangle+|\tilde{0}a  \rangle  \right)/2
.
\end{align}

We now consider two-site system and combine the states $\phi_{i\mu}(a)$ with $\mu=\pm$ for the sites $i=1,2$ to make a pseudo-spin singlet. 
We obtain 
both homogeneous and staggered states, the latter of which is given by
\begin{align}
\phi_{12} (\bm Q)\equiv \frac 1 2 \epsilon_{ab} \epsilon_{\mu\nu} 
\phi_{1\mu}(a)\phi_{2\nu}(b)
,
\end{align}
with the eigenvalue $2\sqrt{10}$ for $\Phi_{1\rm R}-\Phi_{2\rm R}$,
while zero eigenvalue for $\Phi_{1\rm R}+\Phi_{2\rm R}$.
Furthermore the state is the singlet of both channel and pseudo-spin.

In the case (ii) we only briefly sketch the construction.  Let us start from the 
state $\chi_\sigma$ 
described by
\begin{align}
\chi_\sigma = \frac 1{\sqrt 2} \epsilon_{ab} 
c_{a\sigma}^\dagger |b\rangle,
\end{align}
where 
$|b\rangle$ with $b=\pm$ represents the doublet of local $f$-electron states.  
The state $\chi_\sigma$ is channel ($\sigma$) doublet (Cd) and pseudo-spin singlet (Ss).
Application of $\Phi^\dagger$ to $\chi_\sigma$ creates another doublet with $n_c=3$, and further application of $\Phi$ brings back to $\chi_\sigma$.
Thus $\Phi_{\rm R}$ is represented by a $2\times 2$ matrix proportional to $\sigma^x$.
Proceeding in a similar manner as in the case (i), we obtain a staggered CsSs state where $\Phi_{1\rm R}-\Phi_{2\rm R}$ is finite.   Note that inclusion of transfer of electrons between the sites 1,2 mixes the singlet states of (i) and (ii).

In the half-filled case $n_c=2$,  the wave function of the symmetry-broken macroscopic state consists of 
linear combination of the products of local states obtained above.
Without the transfer term, the expectation value of the Hamiltonian is the same in the staggered and homogeneous $\Phi$ states.  With the transfer term, however, only the staggered state is degenerate with the $\Psi^\mu$ order which gains the kinetic energy by homogeneous symmetry breaking.

\subsection{Staggered odd-frequency order}

As in the case of diagonal orders,  the composite off-diagonal order can be regarded as an odd-frequency (OF) order of conduction electrons \cite{abrahams93,emery92}.  
This identification is very useful in deriving the phase transition numerically from the disordered side. Namely, one can search for divergence of the OF response function including only conduction electrons, which is much simpler than treating  the composite operator directly.

We explain more details of the OF order taking
a pairing function with symbolic notation: 
\begin{align}
F_{\mu\nu} (\tau)\equiv -\langle T_\tau c_\mu (\tau) c_\nu 
\rangle
= \psi_0 +\tau\psi_{\rm odd}+ \tau^2\psi_2+\cdots,
\label{pairing}
\end{align}
where $\mu,\nu$ are quantum numbers such as spin, orbital or momentum.
If $\psi_0 =\langle c_\mu c_\nu  \rangle \neq 0$,
we have an ordinary pairing.  
On the other hand, if we have $\psi_0=0$ but  $\psi_{\rm odd} = \langle \dot{c}_\mu c_\nu  \rangle \neq 0$, there is no ordinary pairing, 
but the gauge symmetry is surely broken.
Such state is called the OF pairing in the literature \cite{berezinskii74,balatsky92,emery92}.
The anticommutation property of fermions requires:
\begin{align}
F_{\mu\nu} (\tau) = - F_{\nu\mu} (-\tau).
\end{align}
Without exchange of the indices $\mu,\nu$, however, 
the $\tau$-dependence is in general neither even nor odd.  
For example, it is possible to have 
$\psi_{\rm odd}\psi_2\neq 0$ in Eq.(\ref{pairing}).  
We shall present such an example later in Eq.(\ref{G-tau}).
The crucial point of the exotic pairing is not so much the temporal parity but that
$F_{\mu\nu} (\tau)\neq 0$ with $\tau\neq 0$, even though $\psi_0=0$.
This is evident if one considers systems without inversion symmetry where the spatial parity is not conserved.  Then
the temporal parity is mixed even for simple pairings \cite{tanaka12}.

The fermionic symmetry is made explicit in the frequency space as
\begin{align}
F_{\mu\nu}(z) &= 
\sum_{mn}\frac{\rho_n+\rho_m}{z-E_n+E_m}
(c_\mu )_{mn}
(c_\nu )_{nm} \nonumber\\
& =
-F_{\nu\mu}(-z) 
, \label{pairing-z}
\\
\bar{F}_{\nu\mu}(z) &\equiv 
\sum_{mn}\frac{\rho_n+\rho_m}{z-E_n+E_m}
(c_\nu ^\dagger)_{mn}
(c_\mu ^\dagger)_{nm}\nonumber\\
& = F_{\mu\nu}(z^*)^* 
\label{pairing-bar}
\end{align}
where $\rho_n$ is the statistical weight of each grand-canonical state $n$ which is superposition of different electron numbers. Hence the corresponding energy $E_n$ includes the chemical potential, and $z$ denotes complex frequency appearing through analytic continuation from Matsubara frequencies $i\varepsilon_n$ \cite{kuramoto00}.
Precisely speaking, the gauge-broken states characterized by $\rho_n$ are meaningful only in the thermodynamic limit with the quasi-average \cite{bogoliubov70}.

In analogy to $\Psi^z$, we regard
$\Phi(\bm Q)$
as a part of the corresponding odd-frequency order parameter.  Namely we define
\begin{align}
{\cal O}(\bm Q) &=
\epsilon_{\sigma\rho}
\epsilon_{ab}
\sum_i \exp({\rm i}\bm Q\cdot \bm R_i)
c_{ia\sigma}
\dot{c}_{ib\rho} \nonumber\\
&\equiv {\cal O}_1(\bm Q) +{\cal O}_2(\bm Q),
\label{Phi-L}
\end{align}
where the decomposition is analogous to ${\cal O}^z$ in Eq.(\ref{cal-O}),
as given by
\begin{align}
{\cal O}_1(\bm Q) 
&= 
{\epsilon_{ab}} 
{\epsilon_{\sigma\rho}} 
\sum_{\bm k} \varepsilon_{\bm k}
c_{\bm k a\sigma}  c_{-\bm k-\bm Q\, b\rho},\nonumber\\
{\cal O}_2(\bm Q) &= J\Phi(\bm Q).
 \label{eqn_second_pair}
\end{align}
Note that the staggered nature of 
${\cal O}_1(\bm Q) $ is analogous to the $\eta$-pairing \cite{yang89}.
The differences however are first the presence of orbital degeneracy, and secondly the factor $\varepsilon_{\bm k}$ in the $\bm k$-summation.
The latter implies the finite pairing amplitude at different sites, since $\varepsilon_{\bm k}$ corresponds to the transfer $t_{ij}$ between sites $i,j$
in the site representation.
As in the case of Eq.(\ref{cal_O}), both ${\cal O}_1(\bm Q), {\cal O}_2(\bm Q) $ acquire finite value in the ordered phase.
Hence the on-site composite pairing mixes with an off-site EF pairing \cite{hoshino16}.

Let us introduce the following operators dependent on imaginary times to 
describe possible pairings:
\begin{align}
\displaystyle
O_i^{\rm CsSs}(\tau,\tau') &= 
\epsilon_{ab} \epsilon_{\sigma\rho}
c_{ia\sigma} (\tau) 
c_{ib\rho} (\tau')
, \label{eq_pair_def1} \\
\displaystyle
O_i^{\rm CtSs}(\tau,\tau') &= 
\epsilon_{ab}
c_{ia\sigma} (\tau) 
c_{ib\sigma} (\tau')
, \label{eq_pair_def2} \\
\displaystyle
O_i^{\rm CsSt}(\tau,\tau') &= 
\epsilon_{\sigma\rho}
c_{ia\sigma} (\tau) 
c_{ia\rho} (\tau')
, \label{eq_pair_def3} \\
\displaystyle
O_i^{\rm CtSt}(\tau,\tau') &= 
c_{ia\sigma} (\tau) 
c_{ia\sigma} (\tau')
. \label{eq_pair_def4}
\end{align}
where the triplet (t) has picked up the simplest component ($d_y$) out of the general combinations specified by the $d$-vector \cite{kuramoto00}.

Note that $O_i^{\rm CsSs}(\tau,\tau) = O_i^{\rm CtSt}(\tau,\tau) = 0$ due to the Pauli principle.
In order to calculate susceptibilities, we introduce the two-particle Green function by
$$
\chi^{\rm CS}_{ij} (\tau_1, \tau_2, \tau_3, \tau_4) = \langle T_\tau
O^{\rm CS}_i (-\tau_2, -\tau_1)^\dagger O^{\rm CS}_j (\tau_3, \tau_4)
\rangle
$$
where 
$\rm CS$ represents one of the labels in Eqs.~(\ref{eq_pair_def1}--\ref{eq_pair_def4}).
Using this quantity and the $\tau$-derivatives as in Eq.(\ref{tau_23-derivative}), we define the even-frequency (EF) and OF pairing susceptibilities $\chi_{\bm q}^{\rm CS}$ with 
$\rm CS\rightarrow \rm EF, OF$
by
\begin{align}
\chi_{\bm q}^{\rm EF} &= \frac{1}{\beta} \sum_{nn'} 
\chi^{\rm EF}_{\bm q} (\imu\varepsilon_n, \imu\varepsilon_{n'}) 
, \label{eq_even_suscep}
\\
\chi_{\bm q}^{\rm OF} &= \frac{1}{\beta} \sum_{nn'} g_n g_{n'} 
\chi^{\rm OF}_{\bm q} (\imu\varepsilon_n, \imu\varepsilon_{n'}) 
, \label{eq_odd_suscep}
\end{align}
where $\rm EF$ corresponds to CsSt or CtSs, while
$\rm OF$ to CsSs or CtSt.

For the OF pairing, a form factor 
$g_n = {\rm sgn }\, \varepsilon_n$ 
makes the calculation much easier than using 
$\varepsilon_n$ \cite{jarrell97, anders02, sakai04} 
The EF susceptibility must be positive definite. 
On the other hand, the OF susceptibility given by Eq.~(\ref{eq_odd_suscep}) 
has the additional contribution corresponding to the second term in the RHS of Eq.(\ref{eq_sus2}). 
Then the sign of the OF susceptibility is indefinite.
It seems that the additional contribution was overlooked in Refs.\citen{jarrell97,cox98}, where the sign change of the response function was interpreted as a first-order transition.  
The lowest temperature in the previous calculation is apparently still higher than the instability toward an OF order.

\begin{figure}[t]
\begin{center}
\includegraphics[width=80mm]{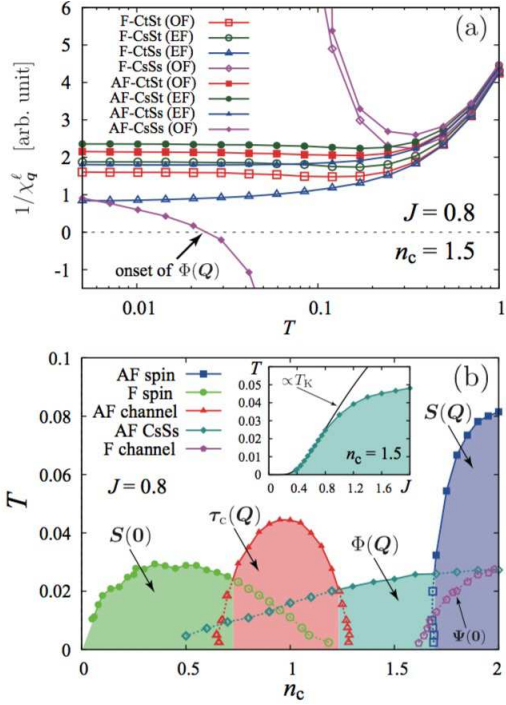}
\caption{
(a) Inverse susceptibilities for EF and OF pairings with uniform (F) or staggered (AF) order.
(b) Phase diagram of the 2chKL 
 in the plane of filling ($n_{\rm c}$) per site and temperature ($T$)
at $J=0.8$.  Here ``AF spin" means the antiferro pseudo-spin (orbital) order, and ``F channel" means the ferromagnetic order, for example.
The inset shows the $J$ dependence of the transition temperature at $n_{\rm c}=1.5$.
The dotted lines in (b) indicate phase boundaries with metastable paramagnetic phase
where another order has already set in at higher temperature \cite{hoshino13}.
}
\label{fig_phase2}
\end{center}
\end{figure}

\begin{figure}[t]
\begin{center}
\includegraphics[width=70mm]{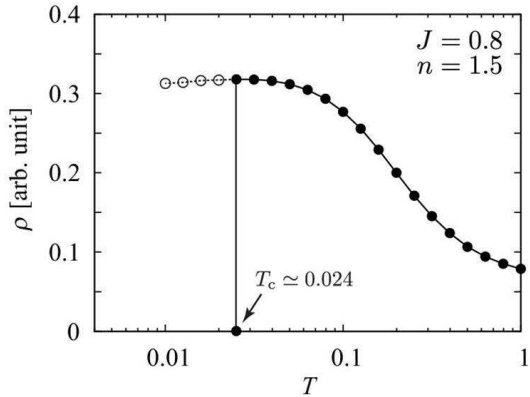}
\caption{
Temperature dependence of the resistivity in 2chKL including the superconducting transition at $T_c$.  At filling $n_c=1.5$, the superconductivity has the highest transition temperature according to Fig.\ref{fig_phase2}.
The empty circles below $T_c$ shows  resistivity in the metastable normal state \cite{hoshino-private}.
}
\label{resistivity}
\end{center}
\end{figure}

Figure~\ref{fig_phase2}(a) shows the temperature dependence of $\chi^{\rm CS}_{\bm q}$ at $J=0.8$ and $n_{\rm c}=1.5$ \cite{hoshino13}.
Here the two ordering vectors $\bm q = \bm 0$ and $\bm q = \bm Q$ are considered, which are called ferro (F) and antiferro (AF), respectively.
Among the eight susceptibilities, only the one with AF-CsSs diverges at $T_{\rm sc} \simeq 0.024$ 
signaling the onset of the staggered OF superconductivity.
Note that the divergence occurs from the negative side as temperature is decreased. 
Since the normal state of the 2chKL is a non-Fermi liquid as seen in electrical resistivity \cite{jarrell96} 
and thermodynamic quantities \cite{hoshino13}, 
the present system becomes superconducting directly from the non-Fermi liquid.
Figure \ref{resistivity} shows a model calculation\cite{hoshino-private} of the resistivity, which is consistent with the original result \cite{jarrell96} except for detection of superconductivity.

Together with the diagonal orders that have been obtained in a similar manner \cite{hoshino13}, the phase diagram of the 2chKL is completed as shown in Fig.~\ref{fig_phase2}(b).
Here the diagonal orders are characterized by the vector operators
\begin{align}
& \hat{\bm S} (\bm q) = \sum_i \hat{\bm S}_i \epn ^{-\imu \bm q \cdot \bm R_i}
, \\
& \bm \tau_{\rm c} (\bm q) =
\sum_{i} \bm \sigma_{\sigma\rho} c^\dagger_{ia\sigma} 
c_{ia\rho}
\epn ^{-\imu \bm q \cdot \bm R_i}
, \\
& \bm \Psi (\bm q) = \frac 12\sum_i
 (\hat{\bm S}_i \cdot \bm \sigma_{ab}) 
\bm \sigma_{\sigma\rho} 
  c^\dagger_{ia\sigma} c_{ib\rho}
\epn ^{-\imu \bm q \cdot \bm R_i}
,
\label{Psi-v}
\end{align}
which describe localized pseudo-spin or orbital ($\hat{\bm S}$), itinerant real-spin or channel ($\bm \tau_{\rm c}$), and composite ($\bm \Psi $) orders, respectively.
The instability toward superconductivity is found in almost all range of the filling unlike the diagonal orders.
This feature seems characteristic of off-diagonal orders as also seen in the attractive Hubbard model \cite{micnas90}.

The inset of Fig.~\ref{fig_phase2}(b) shows the transition temperature
at $n_{\rm c}=1.5$  as a function of $J$, where only the superconductivity is found at $T \geq 0.001$.
Note that the transition temperature for $J\lesssim 0.8$ 
scales with the Kondo temperature $T_{\rm K} \propto \exp [-1/\rho(\mu)J]$ derived in the weak coupling limit.
The value $J=0.8$ was chosen simply for easier calculation, but it turns out that
the value is still in the regime following the weak coupling behavior.

In closing this subsection we mention the result for a related model  called the two-channel, or SU(2)$\times$SU(2),  Anderson lattice \cite{cox98,anders02}.  It has been reported that a homogeneous OF pairing with the CtSt symmetry is realized only in the presence of charge fluctuation of $f$-electrons \cite{anders02}.  
Here the calculation uses the extension of the resolvent method called the NCA \cite{hewson93,kuramoto00}.  It is highly desirable to check the result with more accurate numerical methods such as the CT-QMC.

\subsection{Quasi-particle spectra}\label{sec:qp-spectra}

At half filling, the transition temperatures for the AF-CsSs $[\bm \Phi(\bm Q)]$ and F channel 
$[\bm \Psi(\bm 0)]$ orders are the same within the numerical accuracy as seen in Fig.~\ref{fig_phase2}(b). 
This degeneracy is a consequence of the SO(5) symmetry.
Let us explore further consequence of the symmetry in excitation spectra.
For this purpose we take
the total single-particle spectrum defined by
\begin{align}
A (\bm k,\omega) \equiv 
-\frac 1\pi {\rm Im}\ 
{\rm Tr}\ 
G(\bm k,\omega+\imu\eta),
\end{align}
where the Green function is regarded as a $4\times 4$ matrix, and 
Tr means summation over pseudo-spin ($a$) and channel ($\sigma$) indices.

\begin{figure}
\begin{center}
\includegraphics[width=80mm]{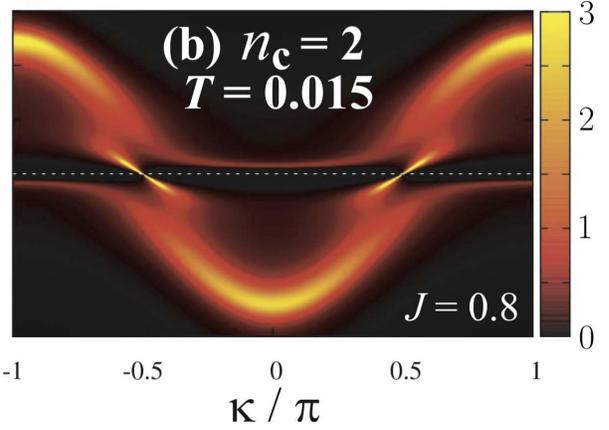}
\caption{
Single-particle spectrum in the composite superconducting state \cite{hoshino-private}.
At half-filling $n_c=2$, the spectrum is the same as 
that of the $\Psi^z$ order, which is shown in Fig.\ref{spectrum_Psi},
provided contributions from each channel are summed.
}
\label{fig_spectrum}
\end{center}
\end{figure}

Figure \ref{fig_spectrum} shows the numerical result in the superconducting phase \cite{hoshino-private}.
Surprisingly, the spectrum looks exactly like the spin sum in the $\Psi^z$ ordered phase shown in Fig.\ref{spectrum_Psi}, even though the nature of the superconducting order is very different from the diagonal composite order.  
To understand the identical spectrum,
we work with the Nambu representation of the Green function with eight components $\nu=1,\ldots,8$:
\begin{align}
\psi_\nu (\bm k)=( c_{a\sigma}(\bm k), 
h_{a\sigma}(\bm k) ), 
\label{8comp}
\end{align}
where $c_{a\sigma}$ corresponds to $\nu\le 4$, while 
$h_{a\sigma}$ corresponds to $\nu\ge 5$.
The latter can be identified as hole annihilation operators
defined by
\begin{align}
h_{a\sigma}(\bm k) = 
{\epsilon_{ab}} 
{\epsilon_{\sigma\rho}} c_{b\rho}(-\bm k+\bm Q)^\dagger . 
\label{hole_k}
\end{align}
We have seen that the PH transformation $P_{\downarrow}$ changes the operator $\Psi^+$ to $\Phi^\dagger$, or equivalently  
$\Psi^-$ to $\Phi$.  
Namely, starting from an ordered state $|\psi^-\rangle$ with
$\langle\psi^-|\Psi^-|\psi^-\rangle \neq 0$, 
we obtain
\begin{align}
\langle \phi_s
|\Phi (\bm Q)|\phi_s\rangle \neq 0, 
\end{align}
where $\phi_s =P_\downarrow \psi^-$ has the same energy as $\psi^-$.
If a physical quantity is invariant under $P_\downarrow$, the expectation value of the quantity is common to the ordered states $\psi_-$ and $\phi_s$.
We now show that the quantity
\begin{align}
\psi_\nu (\tau) \psi_\nu^\dagger
\end{align}
summed over eight components
remains invariant under $P_\downarrow$.  
More generally, any transformation $S$ within the SO(5) symmetry gives
\begin{align}
S\psi_\nu S^{-1} 
= \psi_\mu U_{\mu\nu},
\label{unitary}
\end{align}
with $\{U_{\mu\nu}\}$ being a unitary matrix.  Then summation over the components $\nu$ gives
\begin{align}
S \psi_\nu (\tau) \psi_\nu^\dagger S^{-1} = 
\psi_\nu (\tau) \psi_\nu^\dagger.
\end{align}
Taking $S=P_\downarrow$, in particular, we obtain
\begin{align}
\langle \phi_s| S \psi_\nu (\tau) \psi_\nu^\dagger S^{-1} |\phi_s\rangle =
\langle \psi_-| \psi_\nu (\tau) \psi_\nu^\dagger |\psi_-\rangle,
\end{align}
Hence the Fourier transform Tr\,$G$ is also common to diagonal and off-diagonal orders.  Repeating the same argument for each excited state, we conclude that the spectrum at finite temperature is also identical between $\Phi$ and $\Psi^z$ orders.  

The unitary matrix in Eq.(\ref{unitary}) generates a new Green function matrix associated with the new basis set.  
Hence not only the trace but the determinant det\,$G$ is also common to different order parameters. 
The latter invariance is explored in the next section.

\section{Virtual hybridization}
\label{sec:hybridization}

Let us interpret the electronic spectrum with the composite order more intuitively.
In the case of $\Psi^z$ as shown in Fig.\ref{spectrum_Psi}, 
the quasi-particles consist of two branches: the Fermi liquid branch and the Kondo insulator branch. 
The effective orbital exchange for the up-spin channel goes to zero, while that for the down-spin channel goes to infinity, or vice versa.
Namely, the intermediate value of $J_{\rm eff}$ in the impurity system breaks up into two extreme ones by symmetry breaking,  
which we can use for constructing a fixed point model for the $\Psi^z$ order.   Then, by using the fact that the SO(5) symmetry persists in the total spectrum in the ordered phases, we identify the fixed-point model for the superconducting state.

In the case of $\Psi^z$ order,  the Green function is diagonal with respect to the orbital and spin indices.  In order to compare with the superconducting order $\Phi$, it is convenient to use $\psi_\nu(\bm k)$ in Eq.(\ref{8comp}) as the basis set.   
Namely the subset is taken as the pair
$(c_{a\sigma}(\bm k), h_{a\sigma}(\bm k))$ for each $(a,\sigma)$.
Let us take the channel $\sigma=\downarrow$.  Then
the Green function for each pseudo-spin is given by
\begin{align}
G_\Psi (\bm k, z)^{-1} 
=
\left (
\begin{matrix}
z-\epsilon_{\bm k}-2V^2/z 
& 0 \\
0 & z-\epsilon_{\bm k} \\
\end{matrix}
\right ).
\label{Green2}
\end{align}
The parameter $V$
describes the hybridization with fictitious resonant level that comes from Kondo correlations.
Using the gauge degrees of freedom in the conduction band, it is possible to take $V$ to be real and positive.  
There is no hybridization for the down-spin hole (second row) that 
corresponds to the up-spin electron.
The hole energy becomes the same as the particle energy in the first row by the relation
$-\epsilon_{-\bm k+\bm Q} = \epsilon_{\bm k} $.

For $\sigma=\uparrow$, on the other hand, 
the diagonal elements in the Green function 
are interchanged from those in Eq.(\ref{Green2}).  
In this case the up-spin hole (i.e. down-spin electron) has hybridization.
By arranging the four $2\times 2$ Green functions diagonally, we recover the original $8\times 8$ matrix as the Nambu representation.

The single-particle spectra described by $G_\Psi (\bm k,z)$ have the following three branches:
\begin{align}
E_0(\bm k) = \epsilon_{\bm k}, \quad
E_\pm (\bm k) = \frac 12 (\epsilon_{\bm k}\pm \tilde{\epsilon}_{\bm k}
), 
\end{align}
with 
$\tilde{\epsilon}_{\bm k} \equiv (\epsilon_{\bm k}^2+8V^2)^{1/2}$.

Let us now proceed to the Green function in the CsSs superconducting order $\Phi$ which is characterized by 
$\langle c_{a\sigma}(\bm k)^\dagger h_{a\sigma}(\bm k)\rangle \neq 0$.  
Since neither orbital nor channel symmetry is broken in the CsSs state, we must have  equivalent diagonal elements for different channels in the Green function.
Moreover, by the SO(5) symmetry, 
we should have
\begin{align}
\det G_\Phi(\bm k,z) = 
\det G_\Psi(\bm k,z).
\label{detG}
\end{align}
Thus the Green function of the $\Phi$ order is constrained to the form:
\begin{align}
G_\Phi (\bm k, z)^{-1} 
=
\left (
\begin{matrix}
z-\epsilon_{\bm k}-V^2/z & -e^{i\phi} V^2/z \\
-e^{-i\phi}  V^2/z & z-\epsilon_{\bm k}-V^2/z \\
\end{matrix}
\right ),
\label{Green2a}
\end{align}
where $\phi$ describes the global relative phase between particles and holes, 
which is set to $\phi=0$ in the following.

It is evident that $G_\Phi (\bm k, z)$ is PH symmetric including hybridization.
The off-diagonal part of hybridization indicates that the virtual zero-energy state
hybridize with both particles and holes, which is impossible without gauge-symmetry breaking.
The Hamiltonian of the effective hybridization can be written as
\begin{align}
{\cal H}_{\Phi-\rm hyb} = V\sum_i f_{ia\sigma}^\dagger (
c_{ia\sigma}+h_{ia\sigma}) +{\rm h.c.},
\label{V_hyb}
\end{align}
where 
$f_{ia\sigma}^\dagger$ 
creates a fictitious fermion with zero energy, and $h_{ia\sigma}$ is Fourier-transform of the hole operator given by Eq.(\ref{hole_k}).
Note that the fictitious $f$ states newly acquire the channel degrees of freedom by symmetry breaking to $\Phi$.
Note also that the effective hybridization is smaller by the factor $1/\sqrt 2$ than that in the $\Psi^z$ state where hybridization
works only for a single channel as shown in Eq.(\ref{Green2}). 

In terms of the fixed-point model,
the composite orders are regarded as symmetry-breaking hybridization. 
For example, the order parameter $\Psi^z$  with broken channel symmetry 
can be rewritten as
\begin{align}
\Psi^z = C
\sum_i 
 f_{ia\sigma} ^\dagger \sigma^z_{\sigma\rho} c_{ia\rho} + {\rm h.c.},
\end{align}
with $C$ being a dimensionless constant. 
In the case of $\Phi$-order, we can rewrite 
${\cal O}_2(\bm Q)$ in Eq.(\ref{Phi-L}) as
\begin{align}
{\cal O}_2(\bm Q) = C'
V \sum_{i} 
\left(  
 f_{ia\sigma} ^\dagger +
{\rm e}^{i\bm{Q\cdot R}_i}\epsilon_{ab}
\epsilon_{\sigma\rho} 
f_{ib\rho} 
\right) c_{ia\sigma},
\end{align}
with $C'$ another dimensionless constant.
Here gauge and translational invariances are broken, while
orbital and channel symmetries are preserved.  

As seen from Eq.(\ref{detG}),
the three branches of quasi-particles have the same spectra as those in the diagonal order $\Psi^z$.  
One of the branches with the spectrum $\epsilon_{\bm k}$ corresponds to
the mixture of particles and holes as given by
\begin{align}
d_{a\sigma}(\bm k) = 
\frac 1{\sqrt 2}[
c_{a\sigma}(\bm k)+ 
h_{a\sigma}(\bm k)].
\end{align}
Thus the excitation branch in Fig.\ref{fig_spectrum}, which looks the same as the original conduction band, consists actually of Bogoliubov quasi-particles.
However, the weight of particles and holes is independent of momentum in strong contrast with quasi-particles in BCS superconductors.  
The equal weight of particles and holes means that $d_{a\sigma}(\bm k)$ describes a neutral particle.
On the other hand, the other two branches in Fig.\ref{fig_spectrum} consist of momentum-dependent PH superpositions, together with hybridization with fictitious resonant states.

Let us now derive the time dependence of the anomalous Green function, and analyze the implication of 
Eq.(\ref{pairing}) 
in more detail.
We make the spectral resolution \cite{kuramoto00}
\begin{align}
G_\Phi (\bm k, z) =	
\int_{-\infty}^\infty \frac {d\epsilon }{z-\epsilon} \hat{I}(\bm k,\epsilon)
 ,
\label{Green-I}
\end{align}
where $\hat{I}$ is the spectral intensity matrix given by \cite{hoshino14}
\begin{align}
\hat{I}(\bm k,\epsilon)
 =
\sum_{\nu=0,\pm} 
\left[ 
A_\nu(\bm k)+\sigma^x
B_\nu(\bm k)
 \right] \delta (\epsilon-E_\nu (\bm k))
 ,
\label{Green-spect}
\end{align}
with $\nu = 0,\pm$ specifying the branch of quasi-particles.
Each weight is given by
\begin{align}
& A_0 (\bm k) = 1/2, \quad B_0 (\bm k) = -1/2, 
\label{0-branch}\\
& A_\pm (\bm k) = B_\pm (\bm k) = 
(1\pm \epsilon_{\bm k}/\tilde{\epsilon}_{\bm k})/4.
\label{pm-branch}
\end{align}
The anomalous Green function, which is the off-diagonal element of $G_\Phi (\bm k,z),$
is neither even nor odd function of $z$.
This is to be compared with the odd off-diagonal part in Eq.(\ref{Green2a}) which represents the pair potential in the mean-field theory.

The Green function with the imaginary time $\tau$ is given by \cite{kuramoto00}
\begin{align}
G_\Phi (\bm k, \tau) =	 
\int_{-\infty}^\infty d\epsilon 
\left[ f(\epsilon) -\theta (\tau)  \right]
e^{-\tau\epsilon }
\hat{I}(\bm k,\epsilon),
\label{G-tau}
\end{align} 
where $f(\epsilon)$ is the Fermi distribution function, and $\theta(\tau)$ is the step function.
We consider 
the limit $\tau\rightarrow 0$ in
the off-diagonal element of $G_\Phi (\bm k, \tau)$.
There is no discontinuity at $\tau=0$ since 
\begin{align}
\sum_{\nu=0,\pm} B_\nu(\bm k) =0,
\end{align}
as a consequence of Eqs.(\ref{0-branch}), (\ref{pm-branch}).
However, the anomalous Green function remains finite in the equal-time limit, which looks like a contradiction to OF pairing, 
but can be understood in terms of the mixing with off-site EF pairing \cite{hoshino16}.
Indeed, the local amplitude given by summation over $\bm k$ does vanish.
This is because the off-diagonal element
$
\sum_{\bm k} I_{12}(\bm k,\epsilon)
$
is an even function of $\epsilon$ as a consequence of 
$\epsilon_{-\bm k+\bm Q} = -\epsilon_{\bm k}$.

It is instructive to compare Eq.(\ref{Green2a}) with the Green function of the conventional BCS state.  The latter is given with the basis set
$(c_{\bm k\uparrow}, 
c_{-\bm k\downarrow}^\dagger )$
as
\begin{align}
G_{\rm BCS} (\bm k, z)^{-1} 
=
\left (
\begin{matrix}
z-\epsilon_{\bm k} & -\Delta \\
-\Delta^* & z+\epsilon_{\bm k} \\
\end{matrix}
\right ) \equiv z-\hat{H}(\bm k)
,
\label{BCS}
\end{align}
where $\hat{H}(\bm k)$ is the $\bm k$-component of the mean-field Hamiltonian with the pair potential $\Delta$. 
The particle and hole energies here have opposite signs
in contrast with Eq.(\ref{Green2a}).
This is because the superconducting order is homogeneous, and the PH symmetry 
connects the same energy by the relation 
$
h_{\bm k\uparrow} = c_{-\bm k\downarrow}^\dagger.
$
Since
$\det G_{\rm BCS}(\bm k,z)$
is an even function of $z$,
the off-diagonal part of 
$G_{\rm BCS} (\bm k, z)$
is 
also an even function of $z$.
This is to be contrasted to 
$G_\Phi (\bm k, z)$ given by Eq.(\ref{Green2a}) where 
$\det G_\Phi (\bm k,z)$ 
is neither even nor odd function of $z$.

\section{Relevance to real ${f}$-electron systems}
\label{sec:experiment}

\subsection{Continuation to energy-band picture}

\begin{figure}[b]
\begin{center}
\includegraphics[width=0.99\linewidth]{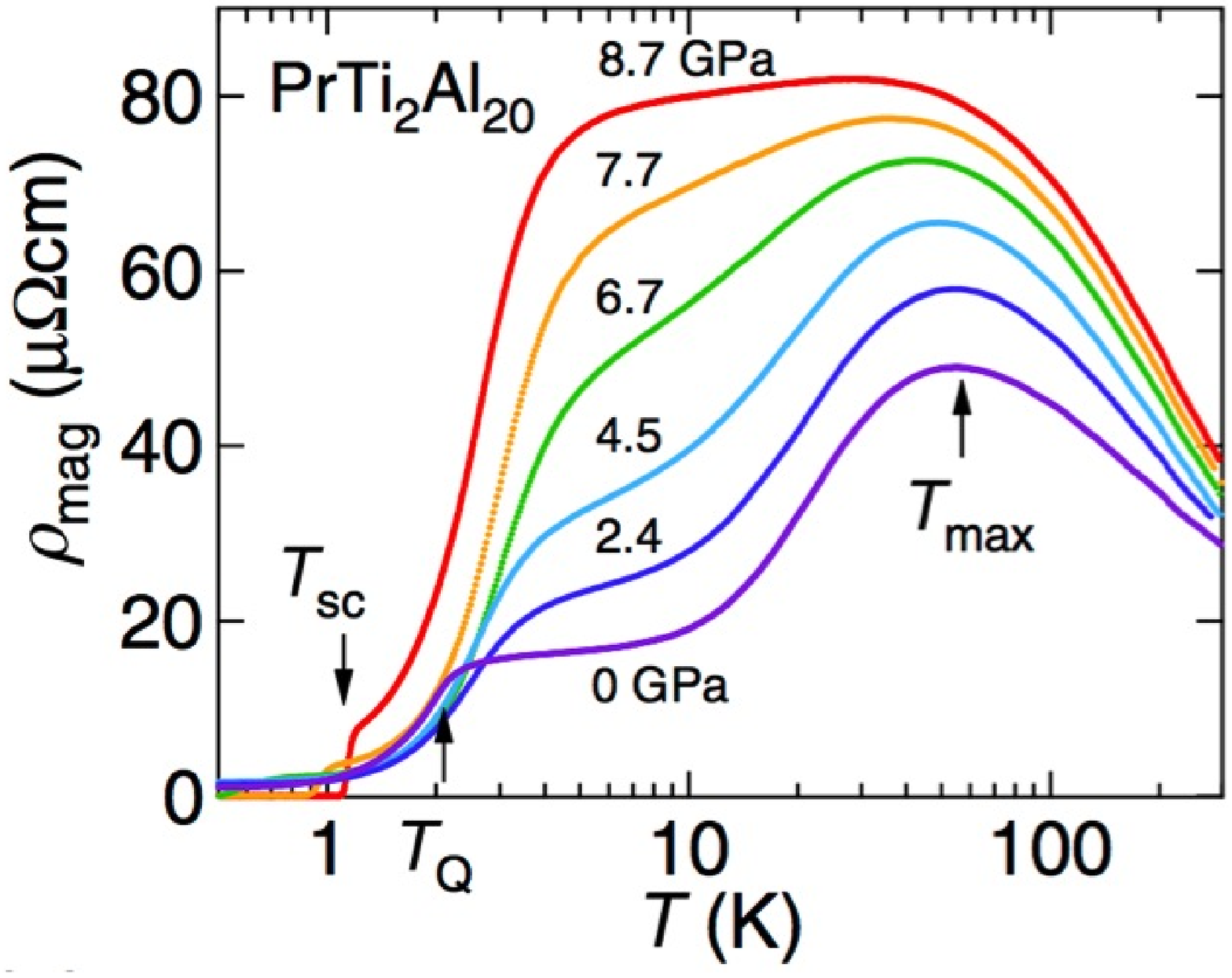}
\includegraphics[width=0.9\linewidth]{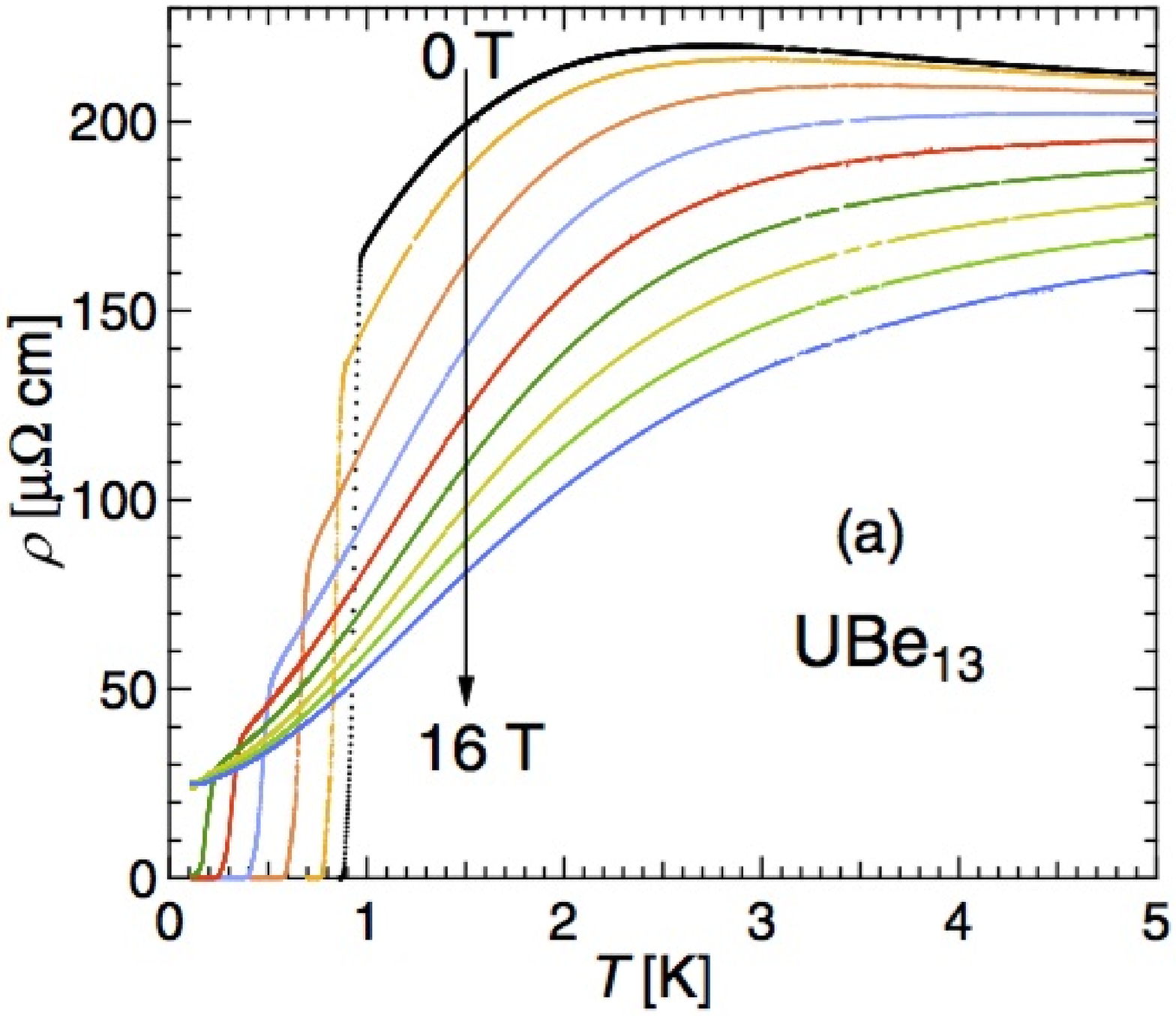}
\end{center}
\caption{
Temperature-dependence of the resistivity in PrTi$_2$Al$_{20}$ under pressure \cite{matsubayashi12} (upper panel), and in UBe$_{13}$ under magnetic fields \cite{shimizu15} (lower panel).
}
\label{rho(T)-exp}
\end{figure}

We discuss possible relevance of results obtained for 2chKL to real physical systems.   
First of all, we have to mention that $f$-electrons in actual materials have 
a wide variety from localized to itinerant characters.  
The 2chKL is applicable only if the charge fluctuation of $f$-electrons is negligible at low energies.  In the opposite limit, the energy-band picture applies.
Then the ground state can either be ordered or paramagnetic, but the non-Fermi liquid never appears as the ground state.  

This restriction of the 2chKL should be kept in mind in understanding the resistivity $\rho (T)$ in the paramagnetic phase.  In the 2chKL, the expected $T$-dependence has been shown in Fig.\ref{resistivity}.  There is no decrease as $T$ goes down to transition temperature $T_c$ of superconductivity.
As shown in Fig.\ref{rho(T)-exp}, however, $\rho(T)$ in actual materials has a decrease slightly above $T_c$.  This is interpreted as a tendency to become a Fermi liquid by charge fluctuations.
Note, however, that 
the Fermi liquid cannot be realized even if charge fluctuations are included in a limited manner, as in the case of the two-channel Anderson lattice \cite{cox98,anders02}.
In the impurity case, namely in 
the two-channel Anderson model, the fixed point is again a non-Fermi liquid for any degree of charge fluctuations.  This result has been
derived by the Bethe ansatz \cite{bolech02}, and by the numerical renormalization group \cite{anders05}.

In the ordered phase, in contrast, we may argue that the composite orders are well described by the 2chKL.  This is because 
the symmetry breaking results in effective hybridization either for a single channel ($\Psi^z$ order) or both channels ($\Phi$ order).
In this way the system escapes from the non-trivial fixed point of the impurity model, and goes to the extremes $J=0$ and/or $J=\infty$.

\subsection{Diagonal orders}
At half-filling of the conduction bands, the staggered orbital order, namely the antiferro quadrupole (AFQ) order,
has the highest transition temperature, as shown in Fig.\ref{fig_phase2}.  
With much lower density of conduction electrons, 
the homogeneous orbital order, namely ferro quadrupole (FQ) order is most stabilized.  
These orders are realized by the orbital version of the RKKY interactions,
and the Kondo effect is not essential.
In actual Pr systems with the doublet CEF ground state, the AFQ is often observed as in PrIr$_2$Zn$_{20}$ \cite{onimaru16}.  On the other hand, PrTi$_2$Al$_{20}$ has a ferro quadrupole (FQ) order at zero pressure\cite{sato12}.  
Interestingly, the entropy of PrV$_2$Al$_{20}$ at the presumed AFQ transition is only $\sim 0.5\ln 2$ \cite{tsujimoto14}, in contrast to the standard value $\ln 2$ as in the case of  PrTi$_2$Al$_{20}$.  It seems that PrV$_2$Al$_{20}$ has stronger hybridization than PrTi$_2$Al$_{20}$, and it is desirable to study the nature of the order in more detail.

\begin{figure}
\begin{center}
\includegraphics[width=0.9\linewidth]{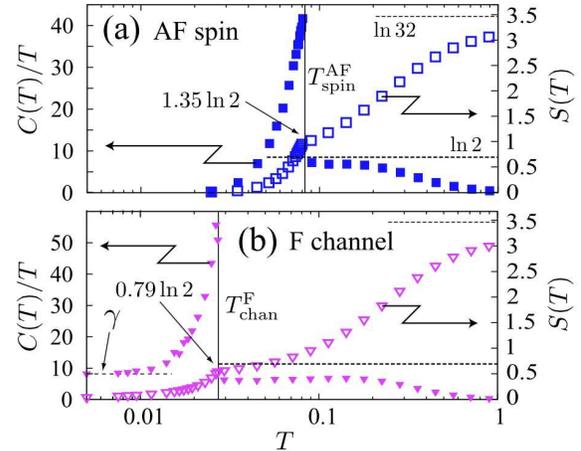}
\caption{
Specific heat and entropy associated with (a) the AF pseudo-spin, and (b) the composite order $\Psi_z$ \cite{hoshino13}.
The entropy includes contribution from conduction electrons.  See text for details.
}
\label{specific-heat}
\end{center}
\end{figure}

Suppose that ordinary AFQ is suppressed by some reason, and the composite $\Psi^z$ order sets in from the paramagnetic phase.  Figure \ref{specific-heat} shows the specific heat and the entropy associated with each transition \cite{hoshino13}.
Numerical calculation gives the entropy at the AFQ transition is about $1.35\ln 2$, while at the $\Psi^z$ transition about $0.79\ln 2$.
Hence, the two diagonal orders may be distinguished by the entropy.
One may naturally ask about the change of entropy associated with the transition.
For this purpose
one can estimate and remove the contribution in the hypothetical disordered state 
below $T^{\rm F}_{\rm chan}$.   
It turns out that $C(T)/T$ remains almost constant in the hypothetical disordered state, and the corresponding entropy amounts to $0.24\ln 2$ 
at $T^{\rm F}_{\rm chan}$ \cite{hoshino13}.   
 Hence the composite order removes the entropy by $(0.79-0.24)\ln 2$ which is close to $0.5 \ln 2$.

\begin{figure}%[b]
\begin{center}
\includegraphics[width=50mm]{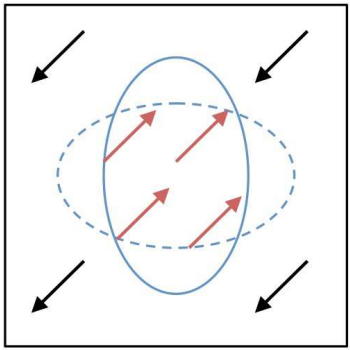}
\caption{
Distributions of real spins (arrows) corresponding to itinerant octupoles where the vector order parameter $\bm \Psi$ points along (110).  The (001) axis is perpendicular to the plane.
The solid and dashed ovals show fluctuating pseudo-spin, i.e., orbital.
}
\label{spin_dist}
\end{center}
\end{figure}

With the homogeneous order $\Psi_z(0)$, the correlation $\langle \hat{\bm S}\cdot \hat{\bm s}_{\sigma}\rangle$ of pseudo-spins (orbitals)
at each site depends on $\sigma$.
As shown in Fig.\ref{fig_illust}(d),
spin-down ($\alpha=2$) conduction electrons make the orbital singlet together with the localized pseudo-spin, while the spin-up ($\alpha=1$) electrons remain essentially free.  
The resultant distribution of each spin in a unit cell should be different 
as illustrated in Fig.\ref{spin_dist}.
Although the difference of the spin distribution is small because it comes from conduction electrons, the deviation of the form factor from the crystalline symmetry may be detected experimentally. 
The resultant anomalous Bragg intensity can  in principle 
be probed by resonant X-ray scattering and spin-polarized neutron scattering.

The most intriguing candidate material to realize the $\bm\Psi$ order is URu$_2$Si$_2$ which has a phase transition called the hidden order (HO) with large anomaly in specific heat.
In spite of great effort of 30 years \cite{mydosh11}, the HO has still escaped identification of its order parameter.
It is clear that the 2chKL model cannot faithfully describe URu$_2$Si$_2$  because the charge fluctuation of $f$ electrons in URu$_2$Si$_2$ is strong enough to form the hybridized Fermi surface, which has been observed by de Haas-van Alphen effect \cite{hassinger10,mydosh11}.
In addition, AF magnetic order appears under pressure \cite{amitsuka07}.
These aspects require more complicated $f$-shell structure than  
described by the 2chKL. 
However, if URu$_2$Si$_2$ indeed
orders with the composite order parameter, the property in the ordered phase may be within the scope of the 2chKL. 
Since the crystalline symmetry of URu$_2$Si$_2$ is tetragonal, the CEF states replacing the non-Kramers doublet is either $\Gamma_5$ doublet \cite{amitsuka94} with dipole moment along the $c$-axis, or two singlets \cite{haule09,kusunose11U}.  If the splitting of singlets is smaller than $T_K$ in the latter case, the orbital Kondo effect can be effective \cite{haule09,toth11} and the composite order may be formed.  However, 
$\Psi^z$ and $\Psi^x, \Psi^y$ are no longer degenerate.  
Figure \ref{spin_dist} illustrates the possible spin distribution where
we have assumed stable directions as $\pm(\Psi^x+ \Psi^y)$.  

With tetragonal symmetry, there should be four equivalent domains composed of $\pm(\Psi^x\pm \Psi^y)$.
With comparable distribution of these domains, the macroscopic symmetry may look like tetragonal.  Even in such cases, the resonant X-ray scattering with E2 (quadrupole) process may be able to detect the octupole order by rotation of the polarization in the scattered light.  However, if the pseudo-spin corresponds to hexadecapoles, the E2 process is still insufficient to probe the composite order that corresponds to triakontadipoles.

Another powerful experimental
method to probe the multipole moment 
is the polarized neutron scattering \cite{kuramoto09,bourdarot14}.  For URu$_2$Si$_2$, induced magnetic form factor has been measured with magnetic field along the $c$-axis,
and some change of the magnetization density has been reported  \cite{bourdarot14}.  
We propose a new experiment where the magnetic field is applied in the $ab$-plane, which may probe 
possible moment distribution illustrated in Fig. \ref{spin_dist}.  Then the magnetic form factor can provide crucial information of the order parameter.

If the HO is homogeneous, the phase transition will be a crossover under external magnetic field that couples with the order parameter.  As a consequence, the sharp anomaly in the specific heat should become rounded with increasing magnetic field  in the $ab$-plane.  
If the broadening of the specific heat is indeed observed, the homogeneous nature of the HO will be established
even though it may not be sufficient to identify the composite order $\Psi$. 

\subsection{Off-diagonal orders}
It is nontrivial whether the off-diagonal order $\Phi$ indeed shows the superconducting property.  The most fundamental is the Meissner effect.  In ordinary superconductors, the phase rigidity against change by 
external magnetic field is the origin of the Meissner effect.
Since the gauge symmetry is spontaneously broken also in composite or OF superconductors, the phase rigidity should be present provided the ordered state is indeed stable against disordered state for small enough magnetic field.

It has been shown by direct calculation of the Meissner kernel \cite{hoshino14} that there is indeed the Meissner effect in the $\Phi$ order in spite of the gapless spectrum.  The magnitude, however, is at most half of the BCS value in the strong-coupling limit.  In the weak coupling case, the magnitude is even reduced.

As shown in Fig. \ref{fig_spectrum}, half of the density of states remains the same as in the normal state.  Hence the Fermi surface keeps the original shape but with the weight reduced to half.  This spectrum is very different from line or point nodes in anisotropic gapless states.  The specific heat behaves in the same way as shown in Fig.\ref{specific-heat} (b).

Another conspicuous signature of the staggered order is emergence of a gapless Nambu-Goldstone mode \cite{hoshino15}.  
In contrast with the homogeneous pairing where
the Anderson-Higgs mechanism leads to the plasmon gap in the Nambu-Goldstone mode \cite{anderson58,nambu60}, the staggered pairing is free from the gap generation.
For the moment, we are not aware of experimental reports of superconductivity that is consistent with this property of the specific heat.

\section{Comparison with related systems}
\label{sec:discussion}

\subsection{Implication to homogeneous OF pairing}

Berezinskii first proposed the 
OF pairing ($s$-wave triplet) as a candidate for the superfluid state in $^3$He with strong repulsion between He atoms \cite{berezinskii74}.  
It has been revived in 1990's, in connection to high temperature superconductivity in cuprates \cite{kirkpatrick91,balatsky92,emery92}.
In most of the literature, the OF superconductivity is assumed to be 
spatially homogeneous.  
However, it has been argued in the weak-coupling limit \cite{heid95} that the OF pairing leads to higher free energy than the normal state.  
Then the ordered state is not stable.  
Furthermore, the Meissner effect in the weak-coupling model
becomes negative \cite{coleman94}.
On the other hand, in composite pairing models \cite{coleman94,abrahams93}, it has been argued that the Meissner effect is small but positive.
There are even arguments \cite{belitz99,solenov09,kusunose11} that cast doubt on the basic fermionic symmetry Eq.(\ref{pairing-bar}),
which is the key to the unstable OF paring \cite{heid95}.
Thus considerable confusion is still present about the homogeneous OF state.
Here we shall briefly comment on the issue in the light of recent theoretical results \cite{hoshino14,otsuki15}.

In most of the previous OF theories, the normal part of the self-energy has been neglected in the spirit of the weak-coupling theory.  However, it has been recognized that the OF pairing requires a coupling constant beyond certain threshold value \cite{abrahams93}.  Hence one should be careful about the validity of the weak-coupling theory, especially concerning the change of free energy near the OF phase transition.

In fact, the Green function in Eq.(\ref{Green2a}) has the normal (diagonal) self-energy as the essential ingredient \cite{hoshino16}.
In addition, the anomalous Green function is {\it not} odd in frequency, although the anomalous self-energy that corresponds to the pairing potential is odd.
The symmetry difference between the anomalous Green function and the anomalous self-energy has not been taken into account in the previous literature.
We suggest that the resonance form $V^2/z$ of the self-energy should not be specific to the 2chKL.  
Recent numerical calculations for single-band models such as
the two-dimensional (2D) Hubbard model \cite{sakai15}
and 2D Kondo lattice \cite{otsuki15}  suggest emergence of resonance or hybridization-type behavior around and below the superconducting transition.
In the 2D Kondo lattice, fluctuations toward the homogeneous OF pairing with singlet $p$-wave are equally strong as the singlet $d$-wave pairing in a special parameter region \cite{otsuki15}.  Namely, the Fermi surface is about to change so as to include the localized spins.
Hence the stability of the homogeneous OF pairing should be analyzed with proper account of the self-energy.

\subsection{SO(5) Symmetry in single-band models}

As we have discussed in Section \ref{sec:off-diagonal}, 
the most spectacular symmetry in the composite orders is 
the degeneracy among the real order parameters;
$\Psi^x,\Psi^y,\Psi^z$ and 
$\Phi_{\rm R}(\bm Q), \Phi_{\rm I}(\bm Q)
$.
The degeneracy is interpreted in terms of the SO(5) symmetry\cite{hoshino-yk14}.
Namely, the order parameters constitute a basis of the vector (or fundamental) representation of the SO(5) group. 
For the two-channel Kondo impurity, presence of an SO(5) symmetry, or equivalently the Sp(4) symmetry,  has long been recognized \cite{affleck92}.  
Probably the most systematic approach is to start from the SO(8) group made up of eight Majorana fermions that originate from free electrons with spin and orbital symmetries \cite{maldacena97}.  Various perturbations break up SO(8) into lower symmetries, but not so far as SU(2)$_{\rm S} \otimes$SU(2)$_{\rm O}$.
Instead of this top-down approach, in this paper we have followed a pedestrian approach that only utilizes the PH symmetry in addition to SU(2)$_{\rm S} \otimes$SU(2)$_{\rm O}$.
A natural question is then how specific is the 2chKL in realizing the SO(5) symmetry.
In the context of high-$T_{\rm c}$ cuprates, 
another kind of SO(5) symmetry has been proposed  \cite{demler-zhang2004} for single-band models such as 
the Hubbard and {\it t-J} models.
In the latter case, the order parameters of antiferromagnetism and the $d$-wave superconductor constitute the approximate basis set of the SO(5) group.

Let us construct the simplest version of single-band models with the exact SO(5) symmetry \cite{scalapino98}.   
We take a pair of sites where the site index $a= 1, 2$ are regarded as an internal degrees of freedom, i.e. a pseudo-spin 1/2.
To distinguish from the spin operator $\bm S_a$ at each site $a$, we introduce a new notation
$\bm T_\sigma$ for the pseudo-spin operator:
\begin{align}
{\bm T}_\sigma = c_{a \sigma}^\dagger \bm\sigma_{a b} c_{b \sigma}/2,
\end{align}
with no summation over $\sigma$.
In the half-filled case with the PH symmetry, the pseudo-spin is conserved under 
$P_\sigma$ given by Eq.(\ref{P_sigma}).  
Hence if a 
model Hamiltonian is isotropic in spin space, and is composed of pseudo-spins only, it should be SO(5) invariant.
Note that SU(2) for the pseudo-spin is not prerequisite to
the SO(5), which is constructed from the (real) spin SU(2) and the PH symmetry \cite{affleck92}.

We take the following Hamiltonian for pseudo-spins:
\begin{align}
{\cal H}
&= 2t (T_\uparrow^x+T_\downarrow^x) +
K_z T_\uparrow^z T_\downarrow^z +
K_2 \left[  
\left(  T_\uparrow^z \right)^2+\left(  T_\downarrow^z \right)^2
\right] \nonumber\\
&+K_\perp \left( T_\uparrow^+ T_\downarrow^-+ T_\uparrow^- T_\downarrow^+ \right).
\label{H_pair}
\end{align}
Here the term with $t$ corresponds to transfer of electrons from one site to the other, which is written alternatively as
\begin{align}
t  (c_{1 \sigma}^\dagger c_{2 \sigma}+c_{2 \sigma}^\dagger c_{1 \sigma})
\end{align}
using the site indices 1,2.
The other terms with $K_j$ in Eq.(\ref{H_pair}) are rewritten 
as
\begin{align}
U\sum_{i=1}^2 \Delta n_{i\uparrow} \Delta n_{i\downarrow}
+V \Delta n_1 \Delta n_2 +J \bm S_1\cdot\bm S_2,
\label{UVJ}
\end{align}
with 
$\Delta n_{i\sigma} =c^\dagger_{i\sigma}c_{i\sigma}-1/2$ and 
$\Delta n_i= n_{i\uparrow}+n_{i\downarrow}-1$
The interaction parameters are related to those in Eq.(\ref{H_pair}) as
$$
U=K_z/4, \ J=K_z/2-K_2, \ V= -K_z/8-K_2/4,
$$ 
which leads to a relation $4(U+V) = J$ as noted in Ref.\citen{scalapino98}.  
The constraint $K_\perp = -K_z/2+K_2$ is necessary to realize the isotropic exchange $J$. 
We assume $J>0$ as a consequence of $U+V>0$.
We remark that Eq.(\ref{H_pair}) does not exhaust the SO(5) invariant terms.  For example, a pair transfer term $T_\uparrow^+ T_\downarrow^+$ has been neglected.

The staggered magnetization along $x$-direction is given by
\begin{align}
S_1^x - S_2^x = \frac 12
c_{a \alpha}^\dagger \sigma^z_{a b} \sigma^x_{\alpha\beta} c_{b \beta}.
\label{S^x}
\end{align}
Combining with another component $S_1^y - S_2^y$,
we construct $S_1^+ - S_2^+ $.
Then the PH transformation $P_\downarrow$ gives
\begin{align}
P_\downarrow \left( S_1^+ - S_2^+  \right) P_\downarrow^{-1} = \frac 12
(\sigma^z \varepsilon)_{a b} \epsilon_{\alpha\beta}
c_{a \alpha}^\dagger c_{b \beta} ^\dagger \equiv \phi^\dagger
,
\end{align}
which creates a pair with spin 
singlet and pseudo-spin triplet.

With these preliminaries we proceed to the symmetry between antiferromagnetism and superconductivity in  
a ladder system where electrons can transfer between the nearest-neighbor cells, each of which is described by Eq.(\ref{H_pair}).  
The new transfer term is invariant under $P_\sigma$ with the staggered phase factor as in Eq.(\ref{PH_site}).
Let us assume antiferromagnetic order where
\begin{align}
S^x(\bm Q)=\sum_j\exp(i\bm Q\cdot\bm R_j) S_j^x,
\end{align}
has a finite average.  
Here $\bm R_j$ runs over all sites, and
the phase factor selects antiparallel spins also in the unit cell of the ladder.  
The PH transformation (\ref{PH_site}) with $i$ being the rung index along the ladder gives 
\begin{align}
P_\downarrow S^x(\bm Q) P_\downarrow ^{-1} = \frac 12\sum_i \left( \phi_i ^\dagger + \phi_i \right)
\equiv \phi_{\rm R}.
\label{PH-ladder}
\end{align}
The RHS of Eq.(\ref{PH-ladder}) represents the order parameter of homogeneous superconductor with spin singlet.
Use of $S^y(\bm Q)$ in Eq.(\ref{PH-ladder}) leads to $-i ( \phi_j ^\dagger - \phi_j )/2 \equiv \phi_{\rm I}$
in the RHS.
Since any direction of the staggered magnetization is energetically equivalent, we recognize
the SO(5) symmetry in the order parameters $\bm S(\bm Q), \phi_{\rm R}, \phi_{\rm I}$
in the half-filled case of the ladder system.

In more standard models such as Hubbard and {\it t-J} models, 
there is no exact SO(5) symmetry. 
However, argument has been advanced \cite{demler-zhang2004} 
that renormalization of bare parameters may drive the system as if the low-energy spectrum has the approximate SO(5) symmetry.
This idea seems motivated by the proximity of antiferromagnetism and $d$-wave superconductivity in cuprates. 
Table I summarizes the SO(5) symmetry aspects comparing between 2chKL and 2D {\it t-J} models.

%%\onecolumn
%\begin{table}
%\begin{center}
%\begin{tabular}{c|c|c}
%\hline
%& 	2 channel Kondo lattice	& 2D {\it t-J} model\\ \hline \hline
%SO(5) symmetry&	exact&	approximate 
%\\ \hline 
%order parameter&	composite (or OF) & 	conventional
%\\ \hline
%diagonal order&  OF spin (channel) & 	spin \\
%& $\bm q=0$&  $\bm q=\bm Q$
%\\ \hline
%off-diagonal order&	OF CsSs
%%channel- \& spin-singlet
%& 	singlet $d$-wave\\
%&  $\bm q=\bm Q$& $\bm q=0$
%\\ \hline
%\end{tabular}
%\caption{Comparison of SO(5) symmetries between 2chKL and the {\it t-J} model. 
%}
%\end{center}
%\end{table}

\begin{figure}%[b]
\begin{center}
\includegraphics[width=90mm]{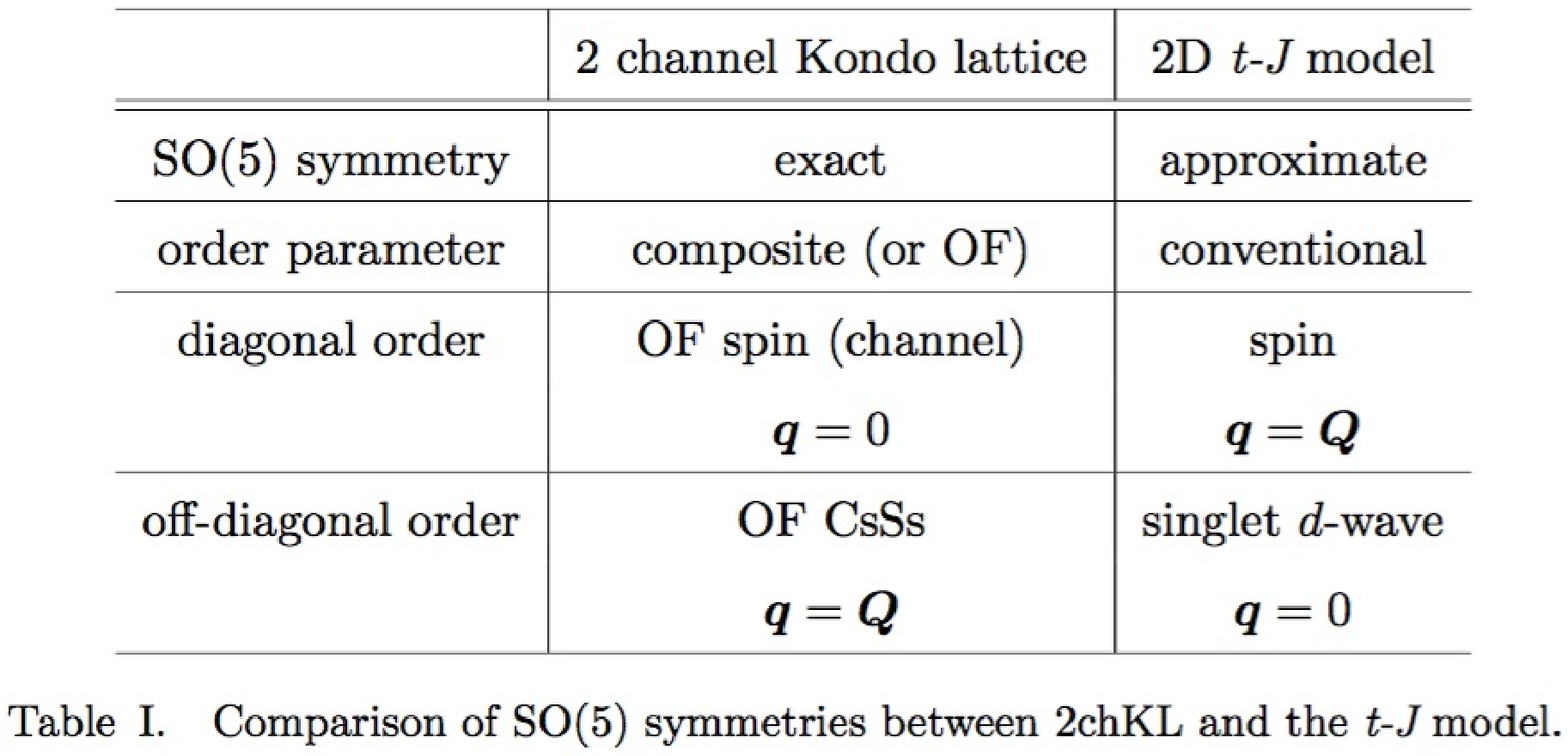}
%\caption{}
\end{center}
\end{figure}

\section{Summary and outlook}

In this paper 
we have described a line of understanding toward exotic orders induced by the orbital Kondo effect.
A special feature is that the finite entropy due to the overscreening Kondo effect can be removed only by electronic orders. 
In other words, the Fermi liquid ground state is impossible in the 2chKL, which is in strong contrast with the ordinary Kondo and Anderson lattices, and with Hubbard and {\it t-J} models.
The special feature of the orbital Kondo effect is shared by the two-channel Anderson lattice where the fractional entropy, $\ln\sqrt 2$ per site, 
remains if any electronic order is absent \cite{bolech02,anders05}.
Note that the fractional entropy at the non-trivial fixed point of the impurity model is most clearly explained in terms of fictitious Majorana fermions \cite{emery92}, which are now under intensive search in condensed matter \cite{alicea12}.
%\textcolor{red}{
In contrast with topological systems where Majorana modes have zero energy and appear at boundaries of systems, 
the present Majorana modes appear in the bulk with macroscopic numbers.
%}
The crucial question is whether these Majorana modes are just like semantics, 
or more like real particles as in fractional quantum Hall systems \cite{laughlin83}.
Although the latter line of understanding is attractive,
more direct signature than the fractional specific heat 
is desired.

It seems useful to further explore the comparative study of SO(5) orders between composite and single-band cases.  An interesting 
problem is to consider alternative kinds of oder parameters.  
For example, 
a unification, under the SO(4) symmetry, 
of triplet superconductivity and antiferromagnetism 
has been proposed 
\cite{podolsky04} for one-dimensional systems such as Bechgard salts. 
In our framework, this corresponds to the case $J<0$ in Eq.(\ref{UVJ}).  
Then the PH transformation of $S_1^x+S_2^x$ in Eq.(\ref{S^x}) 
gives a spin-triplet pair with odd parity for two sites.  The homogeneous triplet superconductivity is degenerate with antiferromagnetism with spin 1 at each rung.

As we have emphasized in Section \ref{sec:experiment},
inclusion of $f$-charge fluctuations is essential 
for making the model more realistic.  
Only with the charge degrees of freedom, the model  is capable of connecting to the energy-band picture.  
For impurity systems, a multi-orbital Anderson-type model has been already studied by the numerical renormalization group \cite{sakai97}. 
Extension of such impurity models to lattice systems may describe more realistic situation of Pr and U systems, especially in the paramagnetic region.
In the ordered phase, on the other hand, we have argued that the charge fluctuations become irrelevant, and the composite orders in the 2chKL provides a useful picture for real systems.  We expect further experimental and theoretical efforts to establish these exotic orders.

\section*{Acknowledgment}
% grant in aid 
The author thanks Shintaro Hoshino for informative discussion and useful comments on the manuscript.  This work was supported by JSPS KAKENHI Grant Numbers 
JP24340072 and JP16K05464.

%.

\end{document}